\documentclass[3p,final,11pt,authoryear]{elsarticle}
\usepackage[colorlinks]{hyperref}
\usepackage[utf8]{inputenc}
\usepackage{graphicx}
\usepackage{mathtools}
\usepackage{amsmath}
\usepackage{lscape}
\usepackage[nameinlink,noabbrev]{cleveref}
\usepackage{enumitem}
\usepackage{natbib}
\usepackage{subcaption}
\usepackage[tableposition = top]{caption}
\usepackage{booktabs}
\usepackage{algorithm}
\usepackage{algorithmic}
\usepackage{hyperref}
\usepackage{float}
\usepackage{tabularx}
\usepackage{adjustbox}
\usepackage{rotating}
\usepackage{lscape}
\usepackage{amsfonts}
\usepackage{graphicx}
\usepackage{amssymb}
\setcitestyle{authoryear,open={(},close={)}}
\usepackage{setspace}

\begin{document}

\begin{frontmatter}

%\title{ Urban Mobility of Immigrantss: Systematic Taste Heterogeneity in Mode Choice Using Mixed Logit Models with Revealed and Stated Preference GPS Panel Data} 

\title{Mobility Behaviour of Immigrants in Canada: Analyzing Mode Choice Using GPS Panel Data and Mixed Logit Models}

    \author[1]{Tareq Alsaleh}\ead{talsaleh@torontomu.ca}
    \author[1]{Bilal Farooq\corref{cor1}}\ead{bilal.farooq@torontomu.ca}
     \author[2]{Zachary Patterson}\ead{Zachary.Patterson@concordia.ca}

    \address[1]{Laboratory of Innovations in Transportation (LiTrans), Toronto Metropolitan University, Toronto, Canada}
    \address[2]{Concordia Institute for Information Systems Engineering (CIISE), Concordia University, Montreal, Canada}
    \cortext[cor1]{Corresponding Author.}

\begin{abstract}
Time spent by Immigrants in Canada, along with their level of integration into society, can influence how they value travel time, cost, and ultimately their mode choices. In this research we examine these relationships using a panel dataset of more than 80,000 trip observations from 100 participants through a custom-built mobile application. A joint revealed preference (RP) and stated preference (SP) framework is used to estimate multinomial logit (MNL) and mixed logit (MXL) models. The level of integration is represented through a composite index capturing economic, social, civic, and health dimensions of integration. Results indicate two distinct patterns. First, the estimated models suggest that new immigrants in the sample exhibit lower sensitivity to in-vehicle travel time than Canadian-born respondents. The mixed logit specification suggests that the value of travel time for the sampled immigrants is approximately 66\% lower than that of Canadian-born residents, with a immigrant-to-Canadian-born ratio of 0.34 that is consistent across both MXL specifications. Second, higher levels of integration are associated with reduced transit use and greater car reliance. A one standard deviation increase in the integration index decreases the probability of choosing public transit by approximately five percentage points. The joint RP-SP specification allows the inclusion of emerging e-mobility alternatives not yet observed in revealed behaviour; these face no inherent preference penalty, competing purely on their level-of-service attributes. Out-of-sample validation using five-fold cross-validation produces mean prediction accuracy between 80\% and 82\% across model specifications. The findings suggest that transit policies in immigrant-receiving cities could prioritize service quality improvements, particularly reductions in access time, which are approximately three times more effective than fare reductions in shifting immigrants toward transit use.
\end{abstract}

\end{frontmatter}

\section{Introduction}
\label{sec:Introduction}

Canada is one of the world's foremost immigrant-receiving nations, with skilled-worker immigration constituting the primary driver of population growth. In the decade preceding this study, the federal government progressively raised annual permanent resident admission targets, reaching approximately 405,000 immigrants in 2021 and setting a trajectory toward 500,000 immigrants per year by 2025 \citep{ircc2022}. As of the 2021 Census, immigrants comprised 23\% of Canada's total population, the highest proportion among G7 nations, and this share is projected to approach 30\% by 2036 \citep{statscan2022}. The metropolitan areas of Toronto, Montr\'{e}al, and Vancouver receive the overwhelming majority of immigrants, though secondary cities are increasingly targeted by provincial nominee programmes to distribute settlement more broadly across regions. 

The scale of the inflow of immigrants has direct implications for urban transportation systems. Cities like Toronto, Melbourne, and New York each absorb hundreds of thousands of immigrants annually, yet relatively little is known about how these individuals make their daily travel decisions. Canada's immigrants are predominantly selected through economic immigration pathways. One would expect they are young, highly educated, skilled, and arrive with strong economic prospects. Their travel behaviour during the critical settlement period, when mode choice habits are still forming, carries long-term consequences for transit ridership, traffic congestion, and urban emissions. Understanding the mode choice patterns of this population is therefore an important area of inquiry for transport planning in immigrant-receiving cities worldwide.
 
For recently landed immigrants, transportation access is not only a logistical convenience but a fundamental prerequisite for social and economic integration. Access to employment, healthcare, education, language training, and social services are all central to the settlement process and dependent on the ability to navigate the urban transportation system \citep{delbosc2023}. Yet the relationship between immigrant status and mobility is bidirectional, while transportation enables integration, the process of integration itself reshapes how immigrants travel. As immigrants acquire language proficiency, obtain driver's licences, secure stable employment, and expand social networks, their transportation mode choices evolve in ways that cannot be captured by static, cross-sectional analyses \citep{welsch2018, asgari2017}.
 
Despite the scale of immigration to Canada and the centrality of transportation to the integration process, urban transportation planning has been slow to recognize immigrants as a population segment with systematically distinct travel behaviour. Conventional travel demand models treat the population as relatively homogeneous with respect to immigrant status, relying on standard sociodemographic segmentation such as income, household size, age, etc, without accounting for the unique constraints and preferences that immigration introduces \citep{harun2021}. This oversight has practical consequences where transit service planning in immigrant-dense suburbs often underestimates demand, while active transportation infrastructure does not account for the different modal preferences observed among recent arrivals \citep{smart2015}.
 
A growing body of empirical evidence demonstrates that immigrants do, in fact, travel differently from the native-born population. Studies across North America, Europe, and other contexts consistently find that immigrants are more reliant on public transit, more likely to carpool, and less likely to drive alone, particularly in the early years following arrival \citep{delbosc2023, asgari2017}. These differences are partly attributable to structural constraints such as lower vehicle ownership, delayed licence acquisition, and income limitations as well as partly to preferences and cultural norms carried from countries of origin \citep{shafi2020, monteiro2021}. Critically, these differences are not permanent: a process of ``transport assimilation'' unfolds over time, with immigrants gradually converging toward host-country travel norms, though the pace and completeness of this convergence vary substantially across individuals and contexts \citep{welsch2025}.
 
Three interrelated limitations in the existing literature motivate this study. First, the dominant analytical approach treats immigration status as a binary variable, i.e. immigrant versus Canadian-born, or, at best, categorizes immigrants by duration of residence or ethnicity. This categorization obscures the multidimensional nature of the integration process, which encompasses economic participation, social connectedness, civic engagement, and health outcomes simultaneously. Second, the data infrastructure underlying most immigrant travel studies relies on cross-sectional household travel surveys or census commute data, which cannot capture within-individual variation over time, seasonal effects, or the full complexity of multi-day travel patterns. Third, while discrete choice modelling has been applied extensively to mode choice in general, it has rarely been adapted to explicitly model the heterogeneity that immigrant status and integration level introduce into travel time and cost sensitivities.
 
This paper addresses these gaps by developing a continuous integration index, a weighted composite of economic (40\%), social (30\%), civic (20\%), and health (10\%) integration dimensions that captures where on the spectrum from newly arrived to fully integrated each immigrant falls. This index replaces the crude binary classification used in prior work and enables systematic investigation of how integration, rather than just time since arrival, moderates transportation mode choice. The analysis draws on approximately 80,000 raw trip observations from 100 individuals in Toronto and Montr\'eal, collected via a GPS-equipped mobile application over several months, supplemented by approximately 622 stated preference scenarios examining hypothetical mode-switching behaviour. By combining mixed logit models with systematic immigrant-driven taste heterogeneity, joint revealed-stated preference estimation, and GPS-based longitudinal panel data, this study offers a methodological and analytical contribution that is, to the best of the author's knowledge, without precedent in the immigrant transportation behaviour literature.

This study makes four specific contributions that collectively address these gaps:
 
\textbf{Contribution 1:} Development and application of a continuous integration index, a weighted composite of economic, social, civic, and health integration that captures where on the spectrum from newly arrived to fully integrated each respondent falls, replacing the binary immigrant/Canadian-born classification.
 
\textbf{Contribution 2:} Specification and estimation of mixed logit models with systematic respondent-driven taste heterogeneity, enabling direct estimation of whether and how respondents differ from Canadian-born population in their sensitivity to travel time and cost, and how this differential evolves with integration level.
 
\textbf{Contribution 3:} Joint RP-SP estimation incorporating an e-mobility alternative not present in respondent current choice sets, with explicit scale parameter estimation and integration as a moderating variable in both RP and SP components.
 
\textbf{Contribution 4:} Use of GPS-equipped smartphone panel data from 100 individuals collected over multiple months in Canadian cities. The panel design prioritizes observational depth over cross-sectional breadth, tracking each individual's full spectrum of daily travel decisions across varying conditions of weather, season, and trip purpose, which provides the first multi-day, individual-level travel behaviour dataset specifically designed to study immigrant mode choice in relation to integration.
 
The remainder of this paper is organized as follows. The Literature Review examines prior work on immigrant travel behaviour, discrete choice modelling approaches in this context, and identifies the research gaps addressed by this study. The Methodology section describes the data collection platform, the GPS-equipped mobile application, the recruitment strategy, and the discrete choice modelling framework comprising multinomial and mixed logit specifications with joint RP-SP estimation. The Results section is structured in three parts: a descriptive analysis of the sample, integration index, and trip diary; model estimation and validation across all four specifications; and behavioural insights including the value of travel time, the integration gradient, and counterfactual modal shift simulations. The Discussion section examines potential implications of the findings for transit planning in immigrant-receiving cities. The paper concludes with a summary of findings, followed by a discussion of limitations and directions for future research.

\section{Literature Review}
\label{sec:litreview}
 
A substantial body of research has examined how immigrants differ from native-born populations in their travel behaviour. Across geographic contexts, the evidence consistently shows that immigrants rely more heavily on public transit, walking, and carpooling than native-born residents, and that this reliance diminishes with duration of residence. In the United States, \citet{chatman2009immigrants} provided an early comprehensive overview of immigrant travel demand, documenting that foreign-born workers were approximately three times more likely to commute by transit and 50\% more likely to carpool than native-born workers, with drive-alone mode shares rising steeply during the first five years after arrival. \citet{asgari2017} confirmed these patterns using the 2009 National Household Travel Survey Florida add-on, finding that foreign-born status significantly increased the probability of choosing transit and non-motorized modes, with recent arrivals showing the strongest effects. \citet{smart2015} demonstrated that immigrant concentration at the neighbourhood level is associated with higher transit and active mode shares, an effect that extends to non-immigrant residents of the same areas, and suggesting that settlement patterns reshape the broader mobility culture of local areas.
 
The concept of transportation assimilation, whereby immigrants gradually adopt the travel patterns of the native-born population, has received sustained empirical attention. \citet{xu2018transportation} revisited this question using repeated cross-sectional U.S. census and American Community Survey data spanning 1980 to 2010, confirming that public transit use declines with length of stay but finding that the rate of assimilation has itself decreased over recent decades. The author attributed this trend to changes in the demographic composition of immigrant cohorts, the geographic distribution of settlement, and the evolving public transit landscape. \citet{lee2021} extended the analysis to emerging shared mobility services, finding that immigrants showed higher predicted bike-share frequency and may be early adopters of certain shared modes.
 
In the Canadian context, \citet{harun2021} documented a systematic positive association between immigrant concentration and transit commute shares across all zones of the Toronto metropolitan area, which revealed a mismatch between transit demand in immigrant settlement areas and suburban service provision. \citet{farber2018transportation} applied the transport and social exclusion framework to Syrian refugees in Durham Region, Ontario, finding that transportation barriers limited participation in social and discretionary activities and negatively affected self-reported wellbeing, including loneliness and sadness. Their mixed-methods approach, combining focus groups, surveys, and GIS-based accessibility analysis, demonstrated that subjective inaccessibility persisted even where objective accessibility measures were adequate, and highlighted the importance of perceived barriers such as language, wayfinding difficulty, and unfamiliarity with transit systems.
 
European and global evidence provide complementary insights. \citet{welsch2018} found that duration of stay was strongly associated with car use among immigrants in Germany, while prior cycling experience in the country of origin significantly predicted bicycle adoption. \citet{welsch2025} showed that after adjusting for socioeconomic and spatial mediators in England, persistent ethnic group differences in bus use remained, which suggests that preferences and cultural factors operate beyond structural constraints. \citet{chen2015} documented the tight connection between socioeconomic status and mode choice for internal immigrants in China, \citet{abulibdeh2023} found nationality to be a significant determinant of airport access mode in Qatar, and \citet{khalil2024} identified perceived transit security as a key factor for immigrants in Egypt.
 
Attitudinal and qualitative research reinforces the behavioural mechanisms underlying these patterns. \citet{shafi2020} found that recent immigrant students in Australia held more positive attitudes toward public transit than native-born peers, with longer-resident showing intermediate attitudes consistent with attitudinal acculturation. \citet{monteiro2021} positioned international residential change as a mobility biography disruption that forces reassessment of travel habits, finding that pre-migration norms and smartphone navigation tools shaped post-arrival mode choices. \citet{garrott2024} identified that modal shift occurs when key events disrupt existing modes and acceptable alternatives become available, a framework directly applicable to the arrival experience of immigrants.
 
Across this evidence base, a consistent methodological limitation emerges: immigrant status is overwhelmingly operationalized as a binary variable or supplemented at best by duration-of-residence categories. No study employs a continuous, multidimensional integration measure capturing economic, social, civic, and health dimensions of the settlement process simultaneously. This categorical treatment obscures within-group heterogeneity and prevents identification of the specific integration pathways that drive mode choice transitions.
  
The multinomial logit model has been the dominant specification in immigrant mode choice research \citep{asgari2017, chen2015, abulibdeh2023}. However, MNL imposes taste homogeneity and the independence of irrelevant alternatives property, both of which are restrictive for heterogeneous populations. Mixed logit relaxes these assumptions by allowing coefficients to vary across individuals according to specified distributions \citep{train2009}, enabling direct estimation of whether immigrants differ from native-born populations in their sensitivity to travel time and cost. Within the immigrant travel literature, no study has specified random parameters logit models with systematic immigrant-driven taste heterogeneity. On the other hand, joint estimation of revealed preference and stated preference data addresses the complementary limitations of each data type: RP data capture actual behaviour but are limited to existing alternatives, while SP data enable evaluation of hypothetical alternatives but are susceptible to measurement noise \citep{brownstone2000}. The foundational framework by \citet{benakiva1990} accommodates the different error variances through a scale parameter, and \citet{bhat2002} demonstrated that ignoring scale differences in joint mixed logit produces biased estimates. Despite the maturity of these methods, no study has applied joint RP-SP estimation to immigrant mode choice with integration as a moderating variable.
 
The foregoing review reveals that while a rich literature examines immigrant mode choice, significant methodological gaps remain. First, migration status is operationalized crudely, typically as a binary variable or duration-of-residence proxy, with no continuous multidimensional integration measure employed \citep{delbosc2023, welsch2025}. Second, heterogeneity-accommodating models such as mixed logit have not been applied to immigrant mode choice to capture migration-driven taste variation \citep{asgari2017, welsch2018}. Third, GPS-based panel data have not been used to study immigrant travel behaviour through discrete choice modelling, despite their demonstrated advantages over cross-sectional methods \citep{patterson2016, harding2021}. Fourth, joint RP-SP estimation has not been applied to immigrant populations with integration as a moderating variable \citep{bhat2002, brownstone2000}. Fifth, the mechanism linking integration, as distinct from mere time since arrival, to mode choice has not been formally modelled \citep{monteiro2021, xu2018transportation}. Table~\ref{tab:litgaps} maps the reviewed immigrant studies against six methodological dimensions, showing that existing work covers mode choice modelling and assimilation effects but that no prior study combines these with a continuous integration measure, taste heterogeneity accommodation, GPS panel data, and stated preference scenarios in a single framework.
 
\begin{table}[htbp]
\centering
\caption{Positioning of the present study relative to existing immigrant travel behaviour literature}
\label{tab:litgaps}
\small
\newcommand{\rhead}[1]{\rotatebox{65}{\parbox{2cm}{\scriptsize\raggedright #1}}}
\begin{tabular}{l c c c c c c}
\toprule
 & \rhead{Mode choice model} & \rhead{Assimilation/ duration effect} & \rhead{Multi-wave or panel data} & \rhead{SP or behav. change} & \rhead{Continuous integration measure} & \rhead{Taste heterogeneity (MXL)} \\
\midrule
\citet{chatman2009immigrants}    &            & \checkmark &            &            &            &            \\
\citet{asgari2017}               & \checkmark & \checkmark &            &            &            &            \\
\citet{smart2015}                & \checkmark &            &            &            &            &            \\
\citet{xu2018transportation}     & \checkmark & \checkmark & \checkmark &            &            &            \\
\citet{harun2021}                & \checkmark &            &            &            &            &            \\
\citet{farber2018transportation} &            &            &            & \checkmark &            &            \\
\citet{welsch2018}               & \checkmark & \checkmark &            &            &            &            \\
\citet{welsch2025}               & \checkmark & \checkmark & \checkmark &            &            &            \\
\citet{chen2015}                 & \checkmark &            &            &            &            &            \\
\citet{shafi2020}                &            & \checkmark &            & \checkmark &            &            \\
\citet{khalil2024}               & \checkmark &            &            &            &            &            \\
\citet{abulibdeh2023}            & \checkmark &            &            & \checkmark &            &            \\
\citet{lee2021}                  & \checkmark &            &            &            &            &            \\
\citet{monteiro2021}             &            &            &            & \checkmark &            &            \\
\midrule
\textbf{This study}              & \checkmark & \checkmark & \checkmark & \checkmark & \checkmark & \checkmark \\
\bottomrule
\end{tabular}
\end{table}
 
\section{Methodology}
\label{sec:methodology}
 
This section describes the data collection process, the post-processing pipeline used to construct the estimation dataset, and the discrete choice modelling framework. %Section~\ref{sec:datacollection} presents the study design, mobile application, recruitment strategy, and survey instruments. Section~\ref{sec:dataprocessing} details the multi-stage data processing workflow that transforms raw GPS traces and survey responses into a modelling-ready dataset with observed and alternative-mode attributes, cost estimates, and contextual variables.
 
\subsection{Data Collection}
\label{sec:datacollection}
\subsubsection{Study Design and Mobile Application}
 
Data collection was conducted between October 2024 and November 2025 in two major Canadian metropolitan areas: the Greater Toronto Area (GTA) and the Greater Montr\'{e}al Area (GMA). These two regions were selected because they collectively receive around 40\% of Canada's immigrant admissions and exhibit substantial variation in transit infrastructure, urban form, and linguistic context, Toronto being predominantly anglophone and Montr\'{e}al predominantly francophone.
 
The study employed a custom-built mobile application (app), \emph{BDMobility}. The detailed technical overview of the app can be found in \cite{alsaleh2025} and \cite{otchere2025}. The app serves three integrated functions: (i)~semi-passive recording of travel trajectories, (ii)~administration of static survey instruments, and (iii)~dynamic generation and presentation of stated preference choice experiments. The semi-passive design leverages the device's GPS and motion sensors through a third-party application programming interface (API) provided by the \textit{MotionTag} platform. This enables automated trip detection, mode inference, and activity recognition, while allowing participants to review, correct, or supplement the recorded information through the app's travel diary interface. This approach balances the data richness of passive GPS tracking with the validation benefits of respondent engagement, addressing the accuracy limitations identified in purely automated systems \citep{harding2021, marra2019, azoulay2024towards}.
 
\begin{figure}[htbp]
\centering
\includegraphics[width=0.95\textwidth]{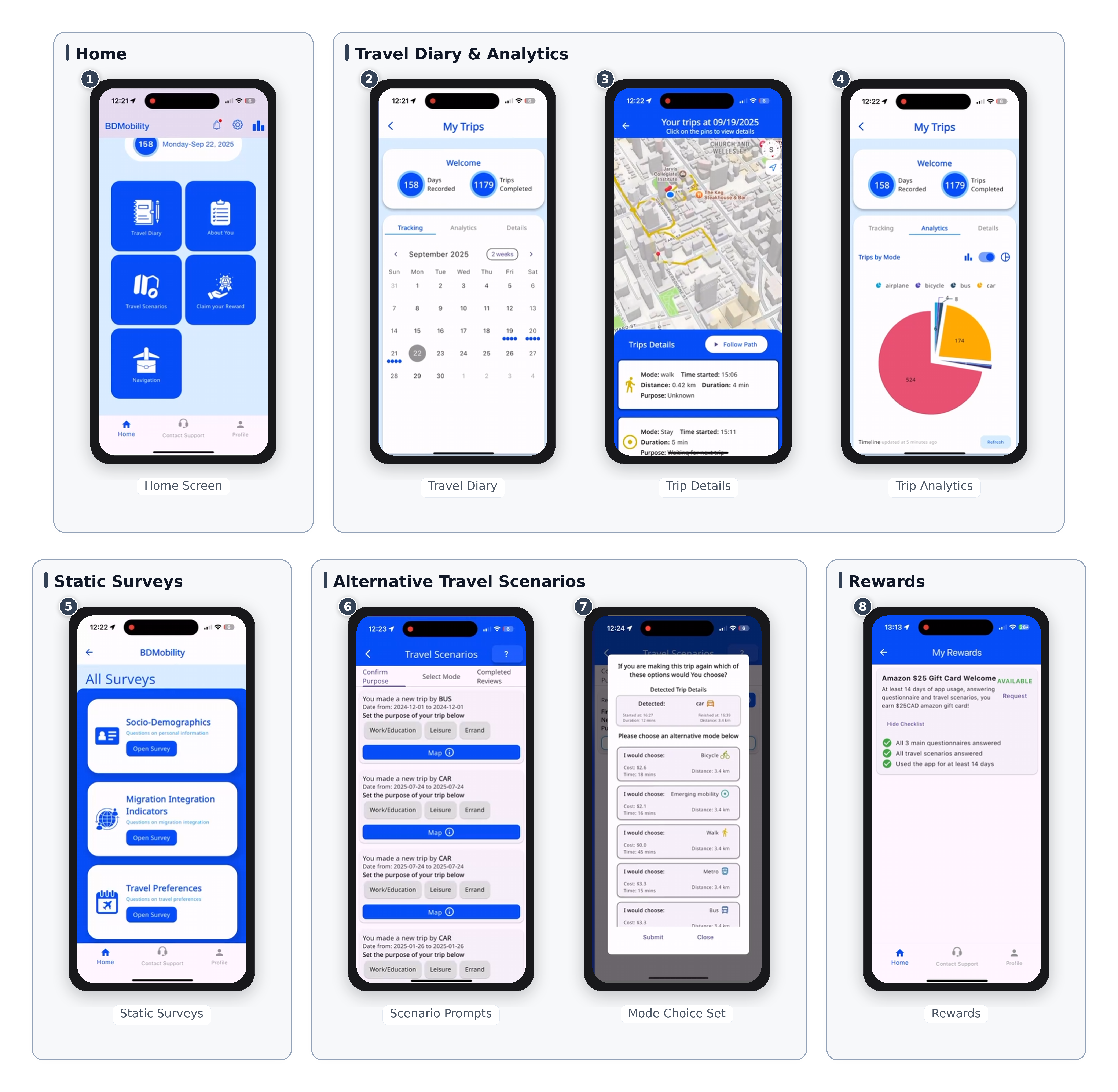}
\caption{BDMobility Application User Interface and Layout}
\label{fig:app}
\end{figure}

The travel diary component, accessible through the app's ``Days'' tab (see Figure~\ref{fig:app}), presents each day's detected trips with recorded attributes including departure and arrival times, trip distance, inferred travel mode, and trip purpose. Participants can modify any of these fields if discrepancies arise between the automated classification and actual trip characteristics; however, the semi-passive design minimizes the need for manual intervention, thereby reducing respondent burden relative to traditional travel diary methods. %Figure~\ref{fig:app} shows the application's user interface layout.
 
\subsubsection{Participant Recruitment}
 
Participants were recruited through a multi-channel strategy combining online campaigns, in-person outreach, and social media dissemination. Online recruitment involved targeted advertisements on platforms frequented by immigrant communities, and university notice boards. In-person campaigns were conducted at community centres,  public transit hubs, and university campuses in both Toronto and Montr\'{e}al. Social media campaigns utilized paid add functions on major social media platforms like TikTok, Instagram and Facebook to reach wider range of participants.
 
To compensate for participation in the multi-day commitment required, each participant received a payment of 25~CAD upon completing a minimum of 14~days of active application usage. This threshold was chosen to ensure sufficient within-individual variation across weekdays and weekends, and to enable the observation of routine and non-routine travel patterns. The 14-day minimum also captures day-to-day variability in mode choice that single-day travel diaries cannot detect.

A deliberate methodological choice underlying this study is the prioritization of longitudinal depth over cross-sectional breadth. Conventional travel surveys, such as the Transportation Tomorrow Survey in the Greater Toronto Area or the Origin--Destination Survey in Montr\'{e}al, collect a single day of travel from tens of thousands of households. While these instruments provide population-representative snapshots, they cannot capture how the same individual's mode choices vary across different days, weather conditions, trip purposes, and times of day. By contrast, the multi-day panel design employed here generates a median of approximately 107 observed trips per participant, recording the full range of daily travel decisions made by each individual over an extended period. This depth of observation per person reveals behavioural patterns that single-day surveys cannot detect: day-to-day mode switching, responses to changing conditions, and the distinction between routine and non-routine travel. The trade-off, well-documented in the smartphone travel survey literature \citep{azoulay2024towards, harding2021}, is a smaller number of unique individuals, which limits the demographic diversity of the sample but does not compromise the identification of the core behavioural relationships under investigation. The objective of this study is not to calibrate a regional demand forecasting model, for which population-representative samples are essential, but to identify the mechanisms through which immigrant status and integration shape mode choice, a task for which observational depth per individual is more informative than cross-sectional coverage.

The recruitment strategy deliberately targeted both immigrants and Canadian-born populations to enable direct comparison within the same data collection framework. First-generation immigrants, second-generation immigrants (Canadian-born with at least one foreign-born parent), and Canadian-born participants with no migration background were all recruited, with Canadian-born participants serving as the control group for investigating how immigrant status and integration level moderate mode choice behaviour.
 
\subsubsection{Survey Instruments}
 
In addition to the passively recorded travel diary, participants completed three sets of questionnaires administered through the application's survey tab, accessible at any point during the study period. All survey questions employed a multiple-choice format with no open-ended items, minimizing completion time and reducing attrition.
 
The \textit{migration integration indicators survey} (37~questions) was designed to operationalize the continuous integration index that is central to this study. The survey instrument draws on the Canadian Index for Measuring Integration (CIMI), a data-driven framework developed to assess immigrant integration outcomes across Canadian municipalities \citep{cimi2020}. The CIMI framework is funded by Immigration, Refugees and Citizenship Canada (IRCC) and has been used since 1991 to benchmark integration outcomes across provinces and census metropolitan areas using data from the Canadian Census, the General Social Survey, and the Canadian Community Health Survey. Its indicator selection was guided by both conceptual and methodological considerations and validated by an Expert Advisory Committee comprising researchers and practitioners in immigrant settlement \citep{cimi2020}. The CIMI examines integration across four dimensions, and our survey operationalizes each through indicators adapted for individual-level measurement: economic integration (employment history, job security, job satisfaction, expertise alignment, perceived wage gap relative to Canadian-born colleagues), social integration (friendship circles, community engagement, sense of belonging at community and national levels, participation in social activities), civic integration (volunteer work, participation in federal and provincial elections, knowledge of the political system), and health and well-being integration (access to healthcare, unmet health needs, daily stress, overall life satisfaction).

Each dimension is assigned a weight in the composite index: economic 40\%, social 30\%, civic 20\%, and health 10\%. These weights are adopted directly from the CIMI's own weighting scheme, which was developed by the CIMI research team and approved by the Expert Advisory Committee. The rationale reflects the relative centrality of each domain to the settlement process: economic integration receives the largest weight because labour market participation is the primary pathway through which immigrants achieve financial independence and establish the routine activity patterns that shape daily travel; social integration captures the interpersonal networks and community ties that influence residential location and trip generation; civic integration reflects institutional engagement that signals deeper embeddedness in the host society; and health integration, while important for overall well-being, operates primarily as a moderating factor in the integration--mobility relationship. Within each dimension, individual indicators are weighted by participants' self-rated importance, ensuring that the index reflects both objective integration outcomes and subjective assessments of their relevance.
 
The \textit{travel preference survey} examined primary transport modes used in the participant's home country for different trip purposes, how these have evolved since arriving in Canada, and the factors that may have influenced modal shifts, including affordability, convenience, safety concerns, and cultural attitudes. This survey also captured the perceived status of different transport modes and the history of mode choice transitions, providing attitudinal context for the revealed preference data.

\subsubsection{Dynamic Stated Preference Scenarios}
 
To investigate willingness to shift between modes, the app incorporated a dynamic stated preference (SP) component that auto-generates alternative travel scenarios based on the participant's observed trip diary \citep{alsaleh2025}. Unlike conventional SP experiments administered as standalone questionnaires, these in-app scenarios are pivoted on actual trip attributes, such as the travel time, waiting time, access time recorded from each participant's own trips, thereby grounding hypothetical choices in real travel experiences and reducing the cognitive burden of evaluating unfamiliar attribute levels.
 
Scenarios were not generated for every recorded trip. Instead, a decision-gate algorithm selected eligible trips based on four criteria: (i)~trip distance (to determine plausible alternative scenarios), (ii)~observed mode choice (distinguishing private, public, and active modes), (iii)~trip frequency (minimizing repeated prompts for similar trips), and (iv)~trip purpose (targeting commute and regular-activity trips). For each eligible trip, the algorithm dynamically formulated alternative-mode attributes using a mechanical calculation derived from observed trip characteristics and price point data from local transit service providers. Cost considerations for active mobility options were tied to the respondent's ownership and accessibility responses from the static survey. Participants were then presented with a choice task: confirm reuse of their observed mode choice or indicate willingness to switch to the presented alternative for a similar future trip.
This design yielded approximately 622~stated preference choice observations across the full sample, providing the SP component of the joint RP- SP estimation.
 
\subsection{Data Processing}
\label{sec:dataprocessing}
 
\begin{figure}[htbp]
\centering
\includegraphics[width=0.8\textwidth]{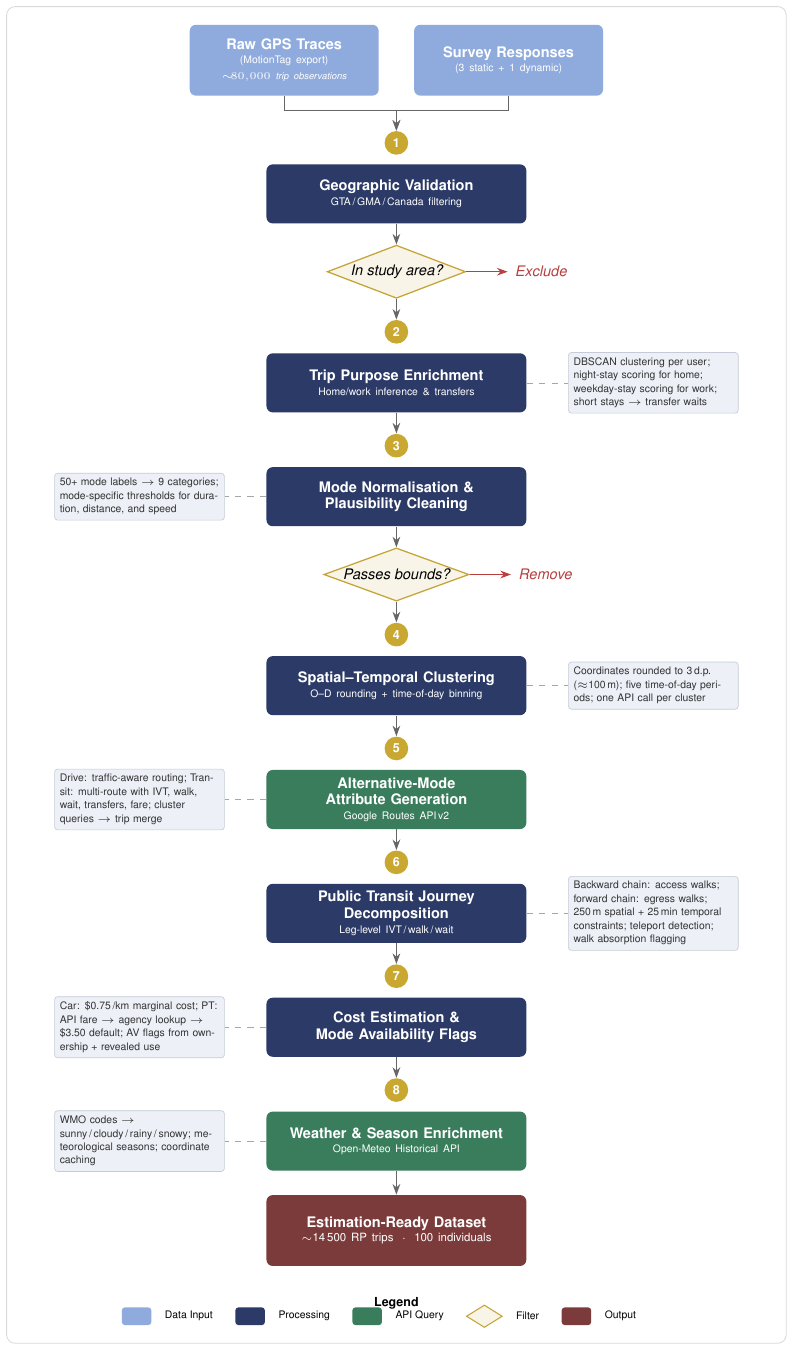}
\caption{Data processing pipeline from raw GPS traces to estimation-ready dataset. Stages are numbered sequentially; diamond nodes indicate quality filters where observations may be removed. Dashed annotations summarize the key operation or external data source at each stage.}
\label{fig:pipeline}
\end{figure}

The raw data produced by the mobile app and survey instruments required extensive post-processing before they could be used for modelling and analysis. A multi-stage pipeline was developed to transform raw GPS traces, travel diary records, and survey responses into a unified, modelling-ready dataset (Figure~\ref{fig:pipeline}). The pipeline encompasses geographic validation, trip purpose enrichment, plausibility-based data cleaning, alternative-mode attribute generation, public transit journey decomposition, cost estimation, and contextual enrichment. %Figure~\ref{fig:pipeline} provides a schematic overview of the pipeline; each stage is described in the subsections that follow.
 
\subsubsection{Geographic Validation and Sample Filtering}
 
Raw GPS trajectories were first parsed from their native format and validated against the geographic boundaries of the study area. Each trip's origin and destination coordinates were tested for membership within two nested spatial envelopes: the Greater Toronto Area, and the Greater Montr\'{e}al Area. Trips falling entirely outside the study area boundaries were excluded, and participants were classified by their primary study area based on the spatial distribution of their recorded trips. This classification was then cross-referenced with survey responses  and filter out-of-scope observations, producing a geographically validated sample linked to the survey data.

A notable data quality challenge arose from the monetary incentive structure of the study, which attracted a substantial volume of fraudulent enrolments. Automated and semi-automated accounts, predominantly originating from outside Canada, attempted to simulate GPS trajectories within the study area to qualify for compensation. These fabricated records exhibited several distinguishing characteristics: implausible spatial patterns such as perfectly straight-line trajectories or instantaneous teleportation between distant locations, unrealistic temporal regularity inconsistent with genuine travel behaviour, duplicate or near-duplicate device identifiers, and survey responses with contradictory or nonsensical demographic information. A multi-stage screening protocol was implemented to identify and remove these records prior to any analysis. First, an email and phone number verification were required, and an IP geolocation and device metadata were cross-referenced against the reported study area. Second, trajectory-level diagnostics flagged trips with physically impossible speeds, zero-variance coordinates, or suspiciously repetitive routing patterns. Third, survey responses were checked for internal consistency, including cross-validation of reported residential location against observed trip origins. All accounts failing any combination of these criteria were permanently excluded. The 100 participants retained in the final analytical sample passed all screening stages, and their trip records were further verified through longitudinal consistency checks confirming plausible day-to-day travel patterns over the multi-month observation period. No suspicious participant entered the modelling dataset. This experience highlights a growing challenge for incentivized mobile-app-based travel surveys and the requirement for robust, multi-layered validation protocols for the deployments of similar data collection platforms.
 
\subsubsection{Trip Purpose Enrichment}
 
The app's automated trip purpose inference was supplemented through a behavioural enrichment procedure applied to the raw GPS event stream. For stationary episodes (stays), a density-based spatial clustering algorithm was applied per participant to identify frequently visited locations from the raw coordinate data. Home and work locations were inferred using a multi-criteria scoring system that evaluated night-time presence duration, number of distinct nights visited, and consistency of being the first or last stop of the day (for home), and weekday daytime duration, number of workdays visited, and regularity of arrival patterns (for work). These anchor locations were then used to validate the inherited purposes from the app's original inference. 
For trips forming part of a public transit journey, the enrichment procedure detected short-duration stays between consecutive transit legs and classified them as transfer waits rather than independent activities, preventing the artificial inflation of trip counts from intermodal connections.
 
\subsubsection{Alternative-Mode Attribute Generation}
 
A central requirement for discrete choice estimation is the availability of attributes not only for the chosen mode but also for the unchosen alternatives in each individual's choice set. To construct these attributes, a spatial-temporal clustering strategy was employed to group trips with similar origins, destinations, and departure times, thereby reducing the volume of routing API queries while preserving sufficient geographic precision.
 
Trip origins and destinations were rounded to three decimal places of latitude and longitude (approximately 100-metre precision), and departure times were binned into five periods (night, AM~peak, midday, PM~peak, evening). Trips sharing the same rounded origin -destination pair and time period were assigned to a common cluster. For each unique cluster, alternative-mode attributes were obtained through the Google Routes API, which was queried for both driving and public transit routing under traffic-aware conditions at representative departure times. The driving query returned total travel time and distance under prevailing traffic conditions. The transit query returned multiple route options, from which the best route for each transit sub-mode (bus, subway/metro, commuter rail) was selected based on total journey time. For each transit route, the API response was decomposed into in-vehicle travel time, walking access and egress time, and total waiting time (computed as the residual between total journey time and the sum of in-vehicle and walking components). The number of transfers, total stops traversed, operating agencies, headway information, and fare data were also extracted where available.
 
The cluster-level alternative attributes were then mapped back to individual trips, producing a dataset in which each observed trip is accompanied by the level-of-service attributes of all feasible alternative modes. Where a participant's observed mode matched one of the queried alternatives, the observed trip's own attributes (travel time, distance) were retained for that mode to preserve revealed preference consistency.
 
\subsubsection{Public Transit Journey Decomposition}
 
For trips where the observed mode was public transit, the raw GPS event stream was analyzed at the individual leg level to decompose the journey into its constituent components. The procedure identified the matched transit track from the raw data, then searched backward in the event sequence to detect access walking legs and platform waiting episodes, and forward to detect egress walking legs. Spatial continuity constraints and temporal plausibility constraints were enforced to prevent spurious linkages. A teleport detection mechanism identified cases where consecutive GPS events showed large spatial discontinuities inconsistent with plausible travel, flagging these for exclusion from access and waiting time attribution.
 
Walking segments identified as access or egress components of a public transit journey were matched to corresponding walk trips in the survey record, and these absorbed walks were flagged to prevent double-counting in the estimation dataset. The decomposition produced, for each observed transit trip, separate estimates of in-vehicle time, access walking time, egress walking time, and waiting time to enable the mode choice models to distinguish between these behaviourally distinct components of the transit journey.

\subsubsection{Cost Estimation and Mode Availability}
 
Trip level cost estimates were computed for each mode in the choice set. Car operating costs were calculated at a marginal rate of \$0.75~CAD per kilometre, reflecting fuel, maintenance, and wear costs without fixed ownership charges (which are sunk costs). Public transit fares were derived from a three-tier hierarchy: (i)~fare data returned directly by the routing API, (ii)~a curated lookup table mapping transit agency names to current fare schedules (covering approximately 60~agencies across the GTA and GMA), and (iii)~a default fare of \$3.50~CAD fallback. For multi-agency transit trips involving transfers, the maximum single-agency fare was applied, reflecting the one-fare transfer policies in effect in both study regions. Walking and cycling costs were set to zero.
 
Mode availability flags were constructed for each trip based on a combination of stated ownership (from the sociodemographic survey), revealed behaviour (whether the participant was observed using that mode), and attribute feasibility. Car availability required either stated car ownership or observed car use, combined with the existence of routed car attributes for that trip's origin-destination pair. Bicycle availability required stated bicycle ownership. Public transit mode availability (bus, subway, train separately) was set to one if the routing API returned a viable route for that sub-mode, and zero otherwise.
 
\subsubsection{Contextual Enrichment}
 
Two contextual variables were appended to each trip record to enable investigation of weather and seasonal effects on mode choice. Trip origin coordinates and dates were used to query the Open-Meteo Historical Weather Archive API, which returned hourly weather codes following the World Meteorological Organization (WMO) classification. These codes were mapped to four categories: sunny (clear sky, mainly clear), cloudy (partly cloudy, overcast, fog), rainy (drizzle, rain, freezing rain, thunderstorm), and snowy (snowfall, snow grains, snow showers). To minimize API query volume, coordinates were rounded to one decimal place (approximately 11-kilometre spatial resolution) and results were cached by unique date and location combinations. Season was assigned using meteorological definitions for the Northern Hemisphere: winter (December-February), spring (March-May), summer (June-August), and fall (September-November).
 
The inclusion of weather and season as contextual variables is particularly relevant for immigrant mode choice analysis, as immigrants from tropical or subtropical climates may exhibit different sensitivity to Canadian winter conditions than the native-born population, potentially moderating the integration and mode choice relationship identified in the discrete choice models.
  
\begin{figure}[htbp]
\centering
\includegraphics[width=0.8\textwidth]{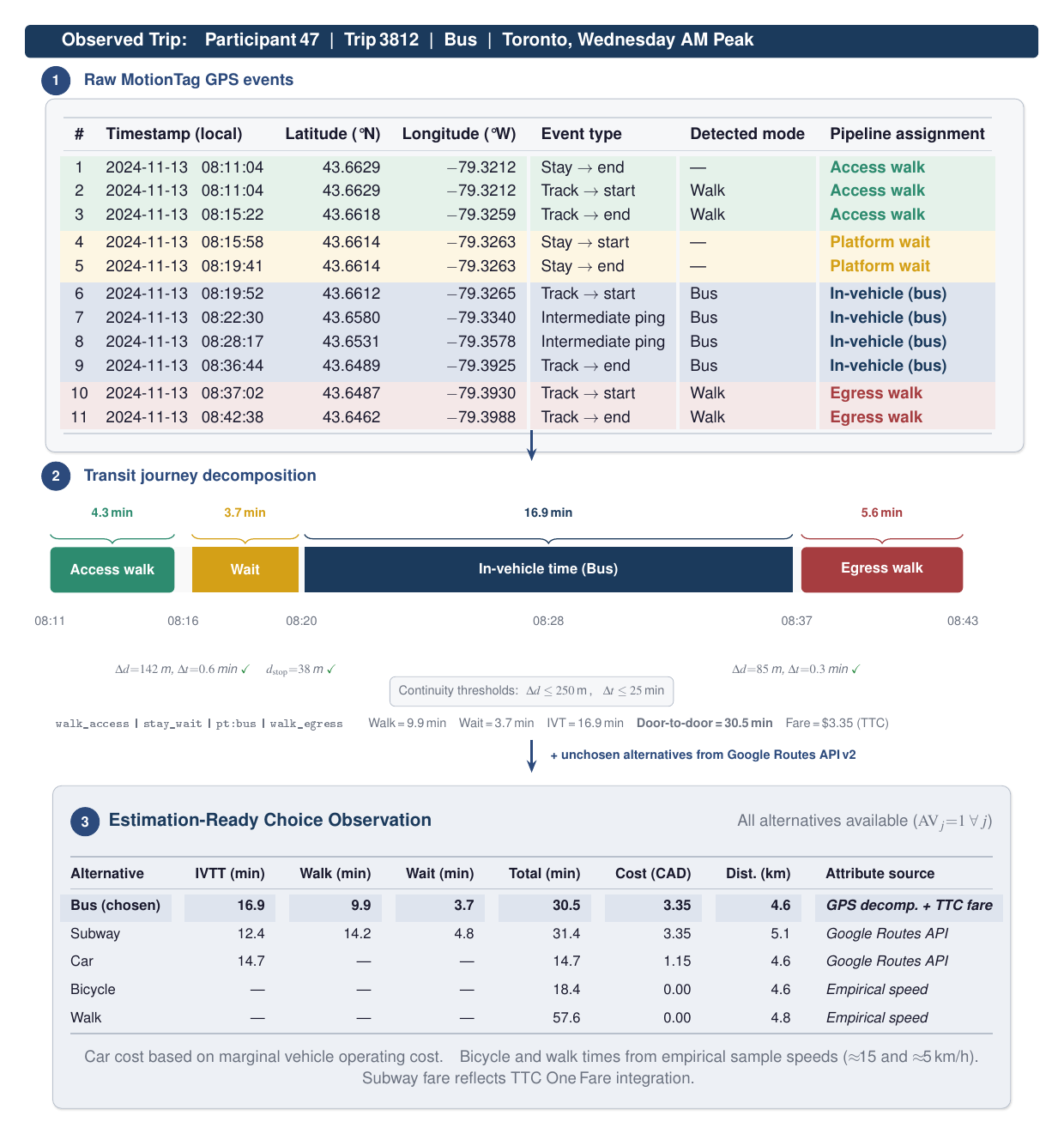}
\caption{Worked example: transit journey decomposition and choice-set generation for a single observed bus trip (Toronto, AM peak). Panel~A shows the raw GPS event stream with pipeline decomposition labels; Panel~B shows the estimation-ready choice set. Chosen-mode attributes are derived from the GPS decomposition; unchosen alternatives are from the Google Routes API. Shading in Panel~A indicates journey phase: access walk, platform wait, in-vehicle time, and egress walk.}
\label{fig:pipeline_example}
\end{figure}

Figure~\ref{fig:pipeline_example} traces a single observed bus trip through the enrichment pipeline. Panel~A shows the raw GPS event stream: the backward search from the boarding event identifies a 4.3-minute access walk and 3.7-minute platform wait, while the forward search identifies a 5.6-minute egress walk, with spatial continuity ($\Delta d \leq 250$\,m) and temporal plausibility ($\Delta t \leq 25$\,min) constraints satisfied at each transition. Panel~B shows the resulting estimation-ready choice set, in which the chosen-mode attributes are derived from this decomposition while unchosen alternatives are generated via the Google Routes API at the spatial--temporal cluster centroid. This process was applied to every trip in the dataset to produce the final estimation-ready observations.

\subsection{Discrete Choice Modelling Framework}\label{sec:modelling}
 
The analysis employs a $2 \times 2$ model comparison framework crossing two dimensions: model structure (multinomial logit versus mixed logit) and data source (revealed preference only versus joint revealed -stated preference). This yields four specifications summarized in Table~\ref{tab:framework}, each addressing a distinct methodological question while maintaining a common core utility specification.
 
\begin{table}[h]
\centering
\caption{Model estimation framework}
\label{tab:framework}
\footnotesize
\begin{tabular}{lcc}
\toprule
 & \textbf{RP Only} & \textbf{Joint RP-SP} \\
\midrule
\textbf{MNL (fixed parameters)} & Model 1 & Model 3 \\
\textbf{MXL (random parameters)} & Model 2 & Model 4 \\
\bottomrule
\end{tabular}
\end{table}
 
\subsubsection{Variable Definitions}
 
Table~\ref{tab:variables} defines the final set of variables entering the utility specifications. Level-of-service variables (cost, time, distance) are mode-specific attributes of each alternative. Sociodemographic and contextual variables are individual-level characteristics that interact with mode-specific constants or enter as taste shifters on selected alternatives.
 
\begin{table}[h]
\centering
\caption{Variable definitions}
\label{tab:variables}
\footnotesize
\begin{tabular}{p{2.8cm} p{5.5cm} p{4.5cm}}
\toprule
\textbf{Variable} & \textbf{Description} & \textbf{Enters utility of} \\
\midrule
\multicolumn{3}{l}{\textit{Level-of-service attributes (continuous, scaled $/10$)}} \\[2pt]
$\text{Cost}_{s}$ & Trip cost (CAD/10): fuel + parking for car; fare for PT & All motorized modes \\
$\text{IVTT}_{s}$ & In-vehicle travel time (min/10) & All modes (incl.\ walk/cycle total time) \\
$\text{Walk}_{s}$ & Walk/access time to stop or station (min/10) & Bus, subway, train \\[4pt]
\multicolumn{3}{l}{\textit{Trip characteristic (continuous, scaled $/1{,}000$)}} \\[2pt]
$\text{Dist}_{s}$ & Trip distance (km/1000) & Mode-specific: car, PT, train, active \\[4pt]
\multicolumn{3}{l}{\textit{Sociodemographic characteristics (binary)}} \\[2pt]
TP\_WS & Trip purpose: work or study ($1$ = yes) & Car \\
CHILD & Has children aged 0--10 ($1$ = yes) & Bus, subway \\
FT & Full-time employed ($1$ = yes) & Subway \\
STU & Student ($1$ = yes) & Train, walk \\
MIG & Foreign-born immigrant ($1$ = yes) & Subway (MNL only) \\
SAFE & Perceives neighbourhood as safe ($1$ = agree) & Subway (M3 only) \\
CYC\_FR & Perceives area as cycling-friendly ($1$ = agree) & Bicycle \\
SNOW & Snow conditions during trip ($1$ = yes) & Walk, bicycle \\[4pt]
\multicolumn{3}{l}{\textit{Integration measure (continuous, mean-centred)}} \\[2pt]
INTEG\_C & Integration index (1--10 scale, centred at $\bar{x}$) & PT modes (all models); active modes (MNL only) \\
\bottomrule
\end{tabular}
\end{table}
 
\subsubsection{Multinomial Logit (MNL)}
 
The multinomial logit model assumes that the random utility $U_{nit}$ of individual $n$ choosing alternative $i$ on choice occasion $t$ is:
\begin{equation}\label{eq:utility}
U_{nit} = V_{nit} + \varepsilon_{nit}
\end{equation}
where $V_{nit}$ is the systematic (deterministic) utility and $\varepsilon_{nit}$ is an i.i.d.\ Type~I extreme value error term. The choice probability follows \citep{mcfadden1972conditional, train2009}:
\begin{equation}\label{eq:mnl}
P(i \mid C_{nt}) = \frac{\exp(V_{nit})}{\sum_{j \in C_{nt}} \exp(V_{njt})}
\end{equation}
where $C_{nt}$ is the set of available alternatives. Car and commuter train serve as reference alternatives ($\alpha = 0$). The systematic utility functions are specified as follows:
 
\begin{align}
V_{\text{car}} ={} & \beta_C \, \text{Cost}_{s} + \beta_T \, \text{IVTT}_{s} + \beta_{D1} \, \text{Dist}_{s} + \beta_W \, \text{TP\_WS} \nonumber \\[6pt]
V_{\text{bus}} ={} & \alpha_{\text{bus}} + \beta_C \, \text{Cost}_{s} + \beta_T \, \text{IVTT}_{s} + \beta_A \, \text{Walk}_{s} \nonumber \\
                   & + \beta_{D2} \, \text{Dist}_{s} + \beta_{Ch} \, \text{CHILD} + \beta_{I1} \, \text{INTEG\_C} \nonumber \\[6pt]
V_{\text{sub}} ={} & \alpha_{\text{sub}} + \beta_C \, \text{Cost}_{s} + \beta_T \, \text{IVTT}_{s} + \beta_A \, \text{Walk}_{s} + \beta_{F} \, \text{FT} \nonumber \\
                   & + \beta_{S} \, \text{SAFE}^{\dagger} + \beta_{Ch} \, \text{CHILD} + \beta_{M} \, \text{MIG} + \beta_{I1} \, \text{INTEG\_C} \nonumber \\[6pt]
V_{\text{train}} ={} & \beta_C \, \text{Cost}_{s} + \beta_T \, \text{IVTT}_{s} + \beta_A \, \text{Walk}_{s} + \beta_{St} \, \text{STU} \nonumber \\
                     & + \beta_{D3} \, \text{Dist}_{s} + \beta_{I1} \, \text{INTEG\_C} \nonumber \\[6pt]
V_{\text{walk}} ={} & \alpha_{\text{walk}} + \beta_T \, \text{TT}_{s} + \beta_{Sw} \, \text{STU} + \beta_{D4} \, \text{Dist}_{s} \nonumber \\
                    & + \beta_{I2} \, \text{INTEG\_C} + \beta_{Sn} \, \text{SNOW} \nonumber \\[6pt]
V_{\text{bike}} ={} & \alpha_{\text{bike}} + \beta_T \, \text{TT}_{s} + \beta_{Cy} \, \text{CYC\_FR} \nonumber \\
                    & + \beta_{D4} \, \text{Dist}_{s} + \beta_{I2} \, \text{INTEG\_C} + \beta_{Sn} \, \text{SNOW} \label{SysU}
\end{align}
 
\noindent where $\beta_C$, $\beta_T$, and $\beta_A$ are generic cost, in-vehicle time, and walk/access time coefficients respectively; $\beta_{D1}$ through $\beta_{D4}$ are mode-group-specific distance parameters; and $\beta_{I1}$, $\beta_{I2}$ are the integration index interactions for public transit and active modes respectively. The superscript $\dagger$ on SAFE indicates that this variable is retained only in M3 (the joint MNL RP-SP specification), where it reached significance at the 5\% level; it was trimmed from M1 following the specification search. The subscript $s$ denotes variables scaled by dividing by 10 (time and cost) or 1{,}000 (distance) to improve numerical stability. Note that the generic time coefficient $\beta_T$ applies to all modes, including total walking time ($\text{TT}_s$) and total cycling time, capturing the disutility of travel time regardless of mode. The integration index enters as a continuous moderator centred at $\bar{x}$: a negative $\beta_{I1}$ implies that higher integration reduces public transit utility relative to car.
 
\subsubsection{Mixed Logit (MXL)}
 
The mixed logit relaxes the MNL assumptions of taste homogeneity and independence of irrelevant alternatives by allowing selected coefficients to vary randomly across individuals \citep{mcfadden2000mixed, train2009}. Two parameters are specified as random:
 
\paragraph{In-vehicle travel time} follows a normal distribution with systematic heterogeneity linked to immigrant status:
\begin{equation}\label{eq:bivtt}
\beta_{T,n} = \bigl(\mu_T + \delta_{\text{MIG}} \cdot \text{MIG}_n\bigr)
  + \sigma_T \cdot z_n, \quad z_n \sim \mathcal{N}(0,1)
\end{equation}
where $\mu_T$ is the population mean, $\delta_{\text{MIG}}$ captures the systematic shift in time sensitivity for immigrants, and $\sigma_T$ captures residual unobserved heterogeneity. A positive $\delta_{\text{MIG}}$ (i.e., less negative mean) indicates that immigrants are less sensitive to travel time. This specification allows the immigrant effect to operate through the \emph{mean} of the random coefficient while preserving individual-level variation.
 
\textit{Cost} follows a negative lognormal distribution to enforce the theoretically correct negative sign \citep{hensher2003mixed}:
\begin{equation}\label{eq:bcost}
\beta_{C,n} = -\exp\bigl(\mu_C + \sigma_C \cdot z_n\bigr),
  \quad z_n \sim \mathcal{N}(0,1)
\end{equation}
 
\noindent The mixed logit choice probability integrates the MNL kernel over the distribution of random parameters:
\begin{equation}\label{eq:mxl}
P_n = \int \prod_{t=1}^{T_n}
  \frac{\exp\bigl(V_{ni^*t}(\boldsymbol{\beta})\bigr)}
       {\sum_{j \in C_{nt}} \exp\bigl(V_{njt}(\boldsymbol{\beta})\bigr)}
  \; f(\boldsymbol{\beta} \mid \boldsymbol{\theta}) \, d\boldsymbol{\beta}
\end{equation}
where $i^*$ denotes the chosen alternative and the product over $T_n$ choice occasions exploits the panel structure. This integral is approximated using 500 Halton draws. To prevent numerical underflow in the panel product for individuals with many trip observations, RP trips are capped per person using stratified sampling (stratified by mode, purpose, and weather conditions) at 300 trips, while retaining all SP observations. All remaining parameters (ASCs, walk time, distance, sociodemographic interactions) are fixed across individuals and maintain the same structure as in the MNL. The MXL specifications are more parsimonious than their MNL counterparts: several sociodemographic interactions that are significant in MNL (including MIG on subway, SAFE, SNOW on active modes, and the active-mode integration interaction $\beta_{I2}$) are subsumed by the random parameters and are therefore excluded from the MXL utility functions.
 
\subsubsection{Joint RP-SP Estimation}
 
Models~3 and~4 extend the framework to joint estimation of revealed and stated preference data following \citet{benakiva1990}. RP and SP observations share a common set of taste parameters but are permitted different error variances through a scale parameter $\mu_{\text{SP}}$:
\begin{equation}\label{eq:scale}
V_{\text{SP}} = \mu_{\text{SP}} \cdot V(\boldsymbol{\beta})
\end{equation}
where $\mu = 1$ for RP observations and $\mu = \mu_{\text{SP}}$ for SP observations. A value $\mu_{\text{SP}} < 1$ indicates that SP responses exhibit greater error variance than RP, consistent with hypothetical bias. The joint models extend the choice set to include an e-mobility alternative (e-scooters and shared electric bikes), available only in certain SP scenarios:
\begin{align}
V_{\text{emob}} ={} & \mu \cdot \bigl(
  \alpha_{\text{emob}}
  + \beta_C \, \text{Cost}_{s}
  + \beta_T \, \text{IVTT}_{s} \nonumber \\
  & + \beta_A \, \text{Walk}_{s}
  + \beta_{D4} \, \text{Dist}_{s}
\bigr) \label{eq:vemob}
\end{align}
 
The initial specification included candidate variables drawn from theory and prior empirical evidence on mode choice, each hypothesized to influence travel behaviour through a specific mechanism such as level-of-service attributes (cost, time, distance) capture rational trade-offs between alternatives; sociodemographic interactions (employment, household structure, car ownership, student status) reflect lifecycle constraints and activity patterns; neighbourhood perception variables (safety, PT-friendliness, cycling-friendliness) operationalize subjective evaluations of the built environment; and the integration index captures the behavioural gradient associated with socioeconomic embeddedness. Each model in the 2$\times$2 framework was estimated independently on the full sample of 100 individuals, and the contribution of each variable was assessed on two criteria: behavioural coherence (consistency of the estimated sign with the hypothesized direction) and statistical reliability. Variables that failed both criteria were removed sequentially, with models re-estimated after each removal. Two classes of variables were exempt from this assessment. Alternative-specific constants were retained regardless of significance, following standard practice in discrete choice modelling \citep{train2009}. The integration index ($\beta_{I1}$) was retained in MXL models despite losing statistical significance, because the absorption of its effect by the random cost and time parameters is an expected property of the mixed logit structure rather than evidence against the integration hypothesis; the MNL models, which lack random parameters, confirm the effect's significance and sign. This process yielded final specifications of 21 parameters for M1, 17 for M2, 24 for M3, and 19 for M4. 
 
All models are estimated on the full sample of 100 individuals using the Apollo package \citep{hess2019apollo} in R. Out-of-sample predictive performance was assessed through five-fold cross-validation: within each individual, RP trip observations are randomly assigned to five folds in round-robin fashion, ensuring that every person contributes observations to every fold. For each fold, all four models are re-estimated on the 80\% training observations, and prediction accuracy is evaluated on the held-out 20\% of RP observations. For MNL models, the predicted mode is the alternative with the highest systematic utility. For MXL models, prediction uses population-mean parameters: $\mu_T$ for in-vehicle time and $-\exp(\mu_C + \tfrac{1}{2}\sigma_C^2)$ for cost. For joint RP-SP models, test evaluation uses RP observations with $\mu = 1$.
 
Value of travel time (VOT) is computed as:
\begin{equation}\label{eq:vot}
\text{VOT} = \frac{\partial V / \partial \text{Time}}
                  {\partial V / \partial \text{Cost}} \times 60
\quad \text{(CAD/hr)}
\end{equation}
For MNL, this reduces to the ratio $\beta_T / \beta_C \times 60$. For MXL, individual-level conditional estimates are obtained via Bayes' rule \citep{revelt1998mixed, train2009}, recovering the posterior mean of each person's random parameters given their observed choice sequence. This avoids the inflation that arises from unconditional simulation when the cost distribution has heavy tails.
 
\section{Results}\label{sec:results}
 
This section presents the empirical findings in three parts. Section~\ref{sec:descriptive} characterizes the sample, the immigrant integration index, the summary of recorded trip diaries, and the stated preference data. Section~\ref{sec:estimation} reports the estimated model parameters, goodness-of-fit measures, and cross-validation results across the four-model framework. Section~\ref{sec:behavioural} interprets the parameters in behavioural terms, including the immigrant time sensitivity differential, the integration gradient, and counterfactual modal shift simulations.
 
\subsection{Descriptive Analysis}\label{sec:descriptive}
 
The final sample comprises 100 individuals who participated in a mobile app panel deployment across the Greater Toronto and Greater Montreal metropolitan areas. The app continuously recorded GPS-based trip observations, yielding approximately 80{,}000 raw trip records. After a processing pipeline that removed incomplete or noise records, imputed missing attributes, joined records for the same trip, and reassigned modal attributes from network data, approximately 14{,}500 revealed preference (RP) trips and 622 stated preference (SP) scenarios remained for model estimation.Figure~\ref{fig:spatial} presents the spatial distribution of recorded trip origins across the two study areas, illustrating the coverage of the GPS panel across the Greater Toronto and Greater Montr\'{e}al metropolitan areas.

\begin{figure}[htbp]
\centering
\includegraphics[width=\textwidth]{Fig_SpatialDensity.png}
\caption{Spatial distribution of recorded trip origins across the
two study areas: (a) Greater Toronto Area and
(b) Greater Montr\'{e}al Area. Colour intensity reflects trip
origin density.}
\label{fig:spatial}
\end{figure}

\begin{figure}[htbp]
\centering
\begin{subfigure}[b]{0.32\textwidth}\centering\includegraphics[width=\textwidth]{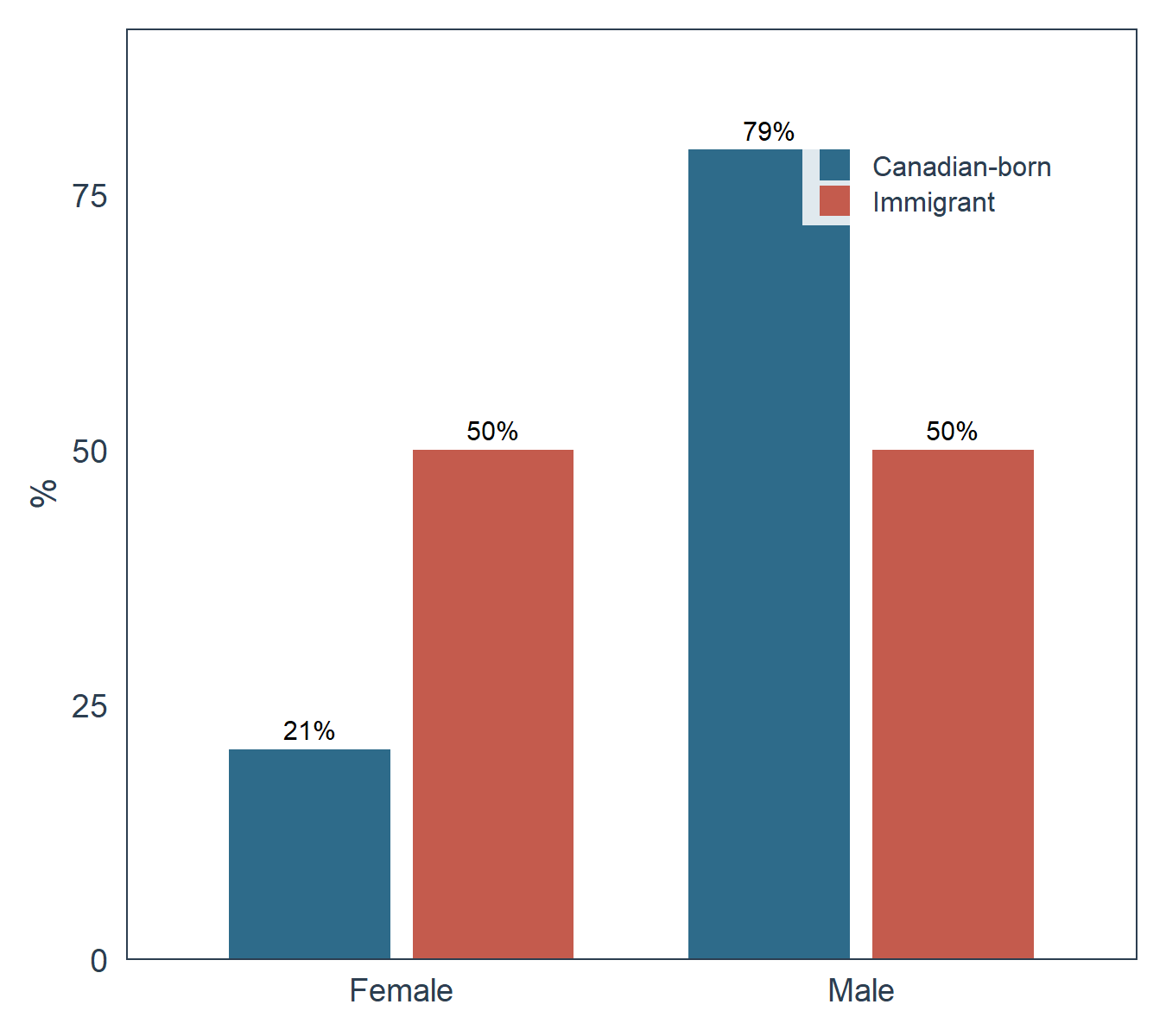}\caption{Gender}\label{fig:gender}\end{subfigure}\hfill
\begin{subfigure}[b]{0.32\textwidth}\centering\includegraphics[width=\textwidth]{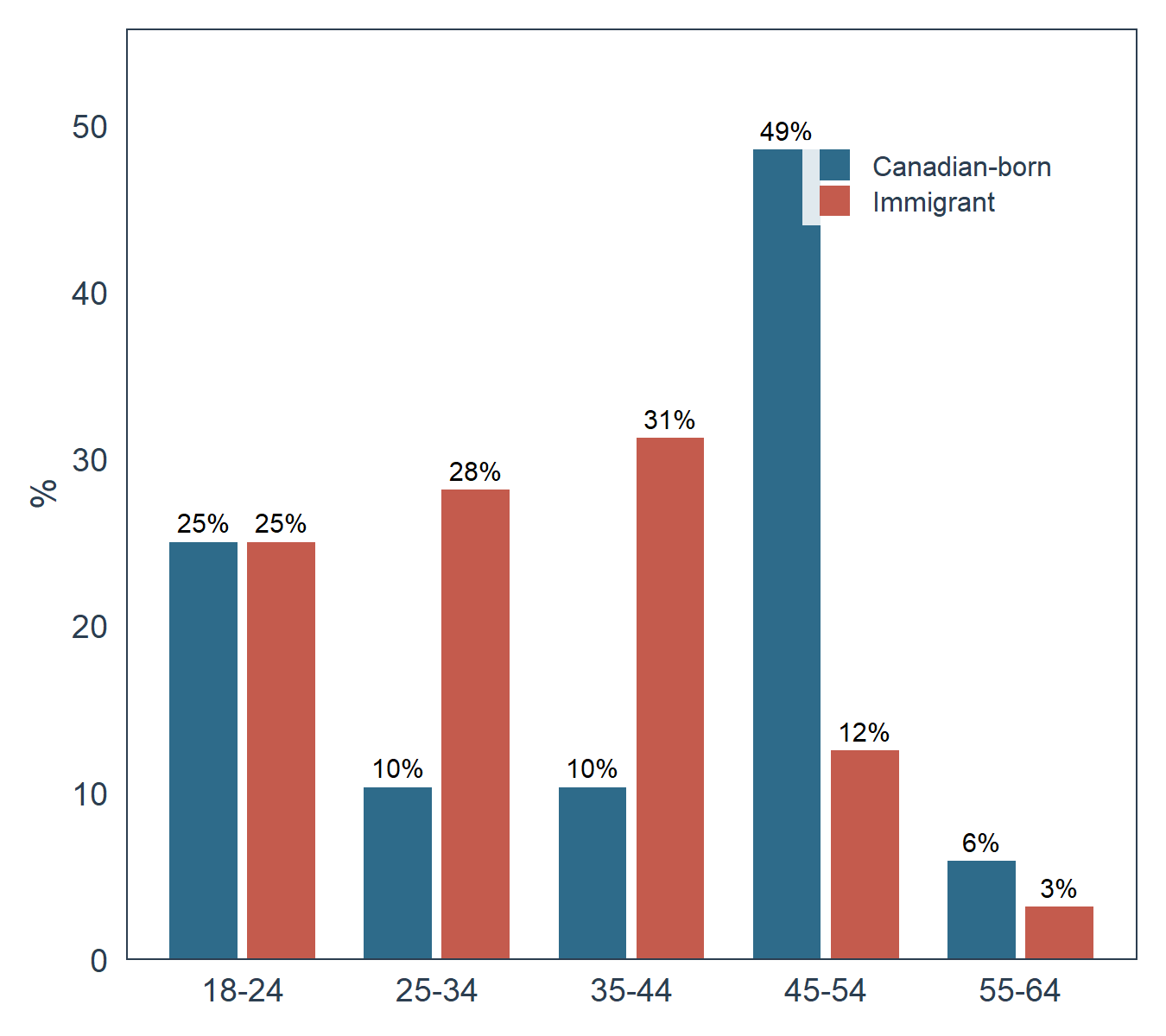}\caption{Age}\label{fig:age}\end{subfigure}\hfill
\begin{subfigure}[b]{0.32\textwidth}\centering\includegraphics[width=\textwidth]{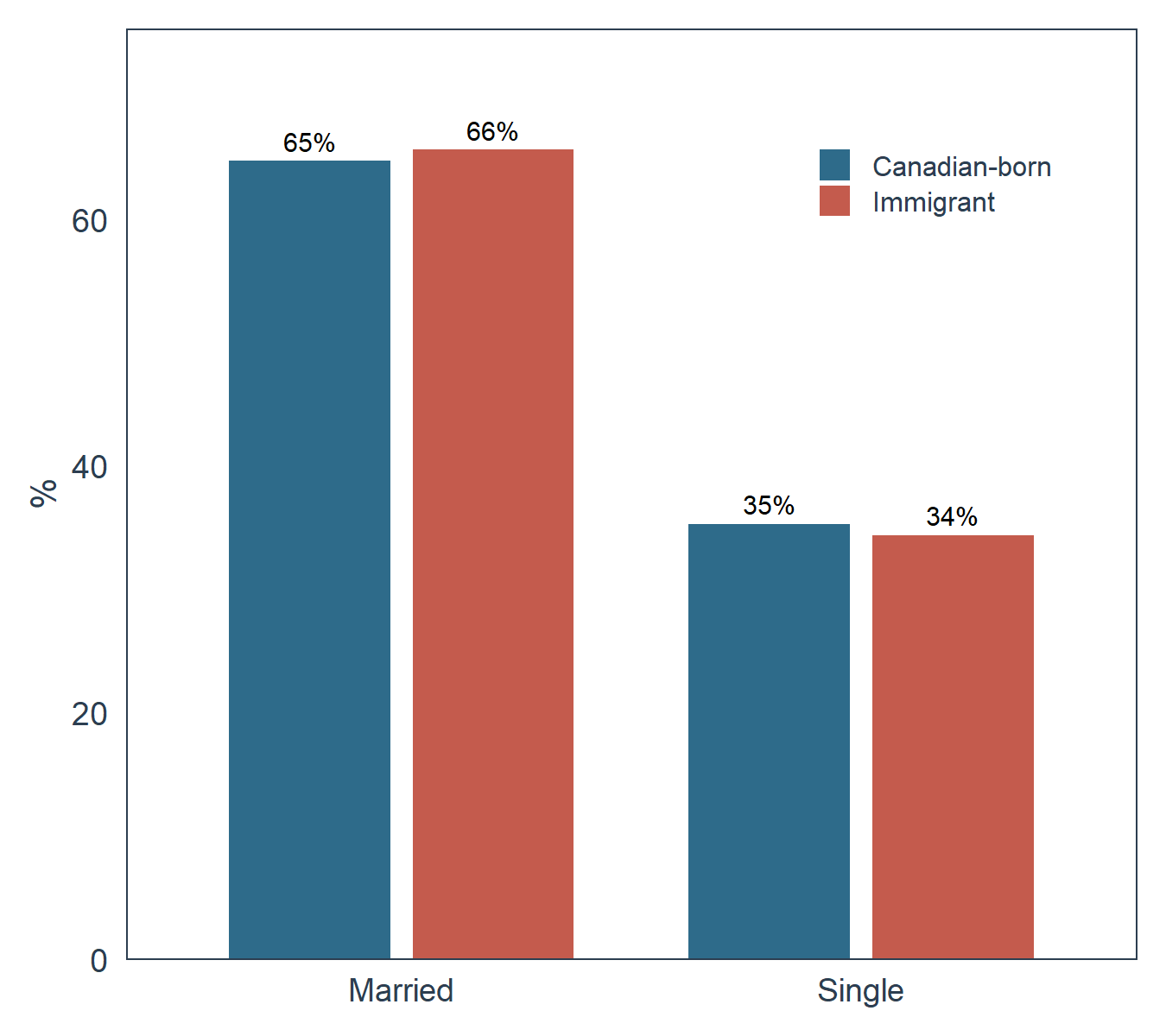}\caption{Marital status}\label{fig:marital}\end{subfigure}\\[6pt]
\begin{subfigure}[b]{0.32\textwidth}\centering\includegraphics[width=\textwidth]{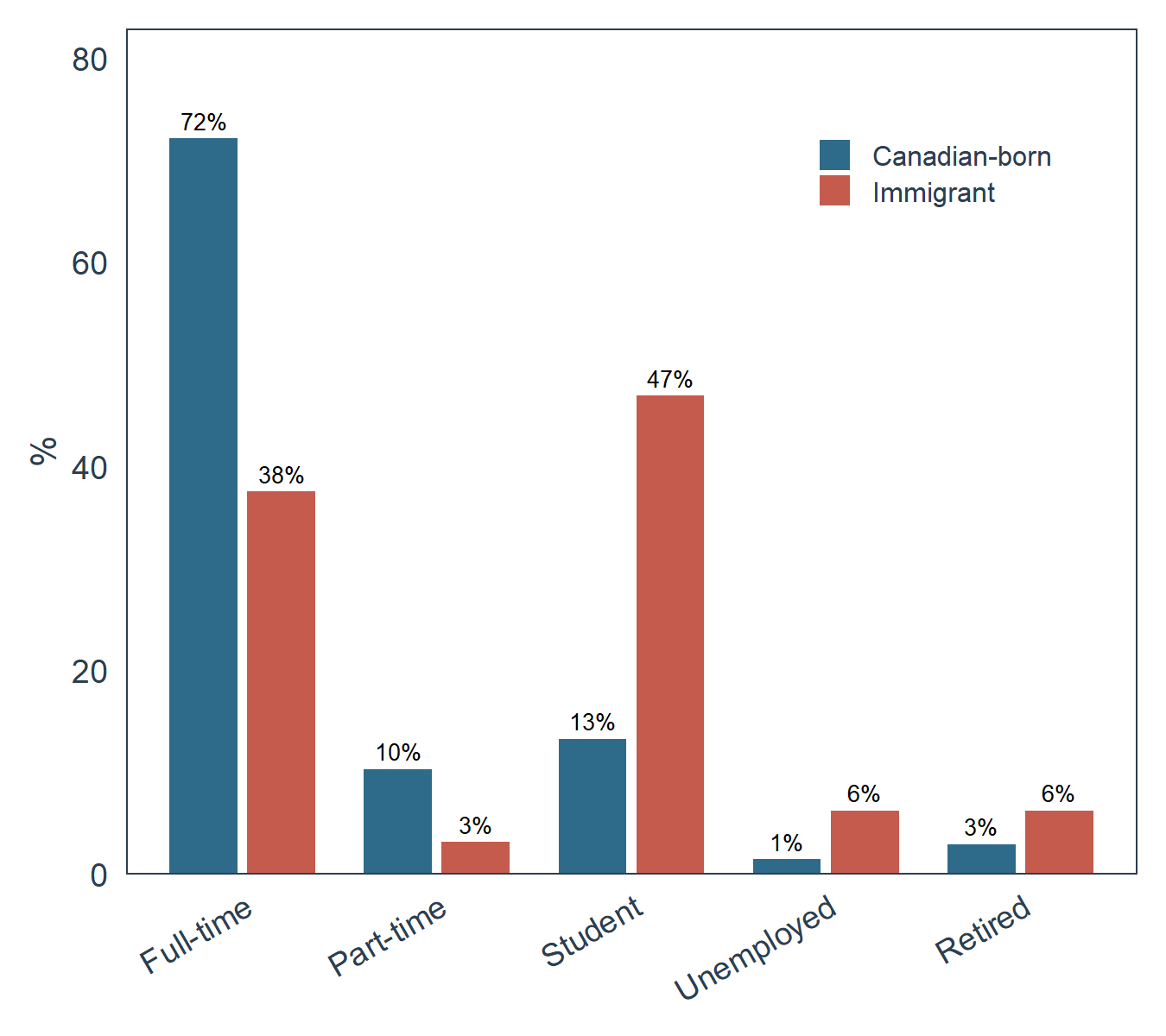}\caption{Employment}\label{fig:employment}\end{subfigure}\hfill
\begin{subfigure}[b]{0.32\textwidth}\centering\includegraphics[width=\textwidth]{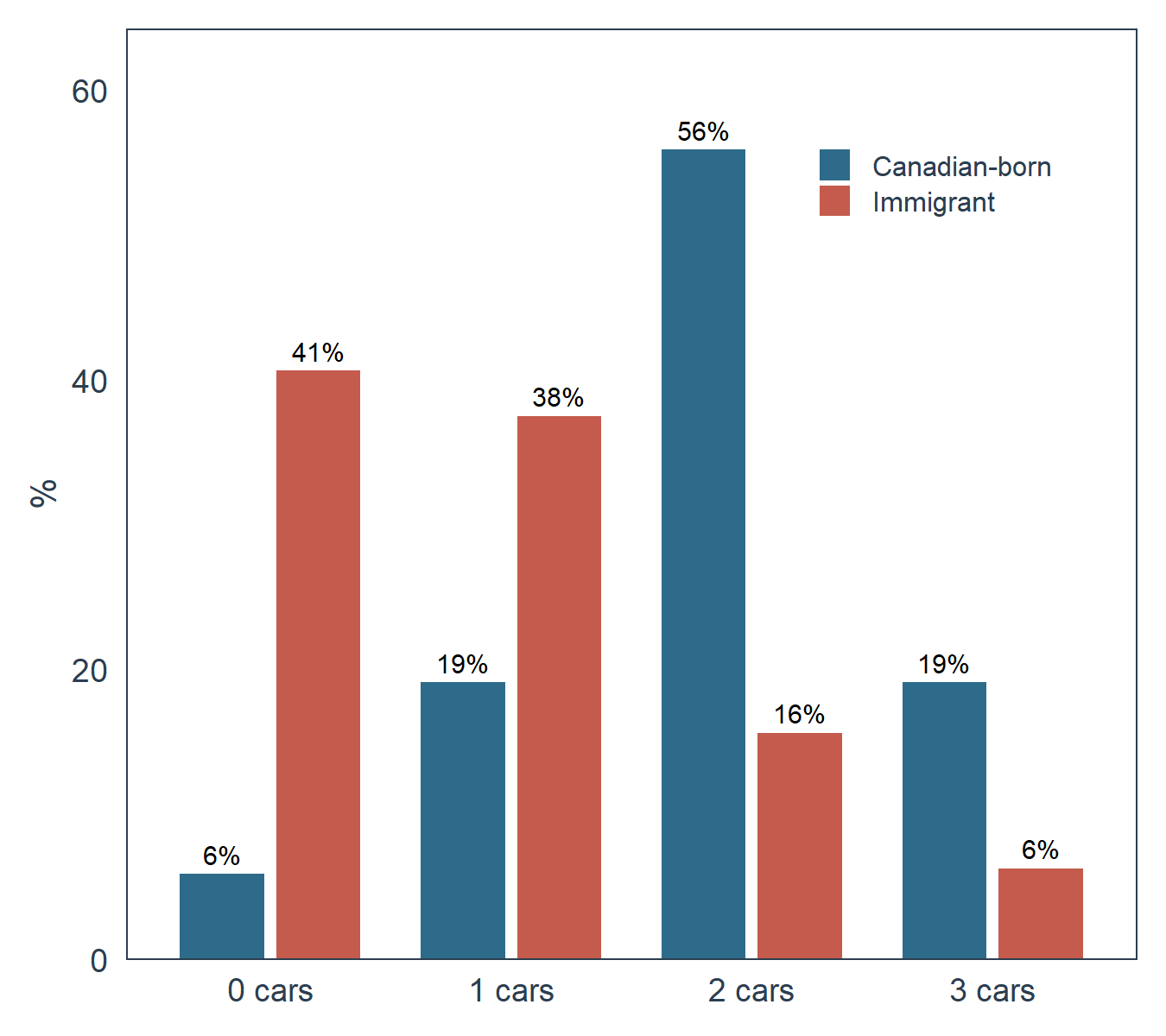}\caption{Car ownership}\label{fig:carown}\end{subfigure}\hfill
\begin{subfigure}[b]{0.32\textwidth}\centering\includegraphics[width=\textwidth]{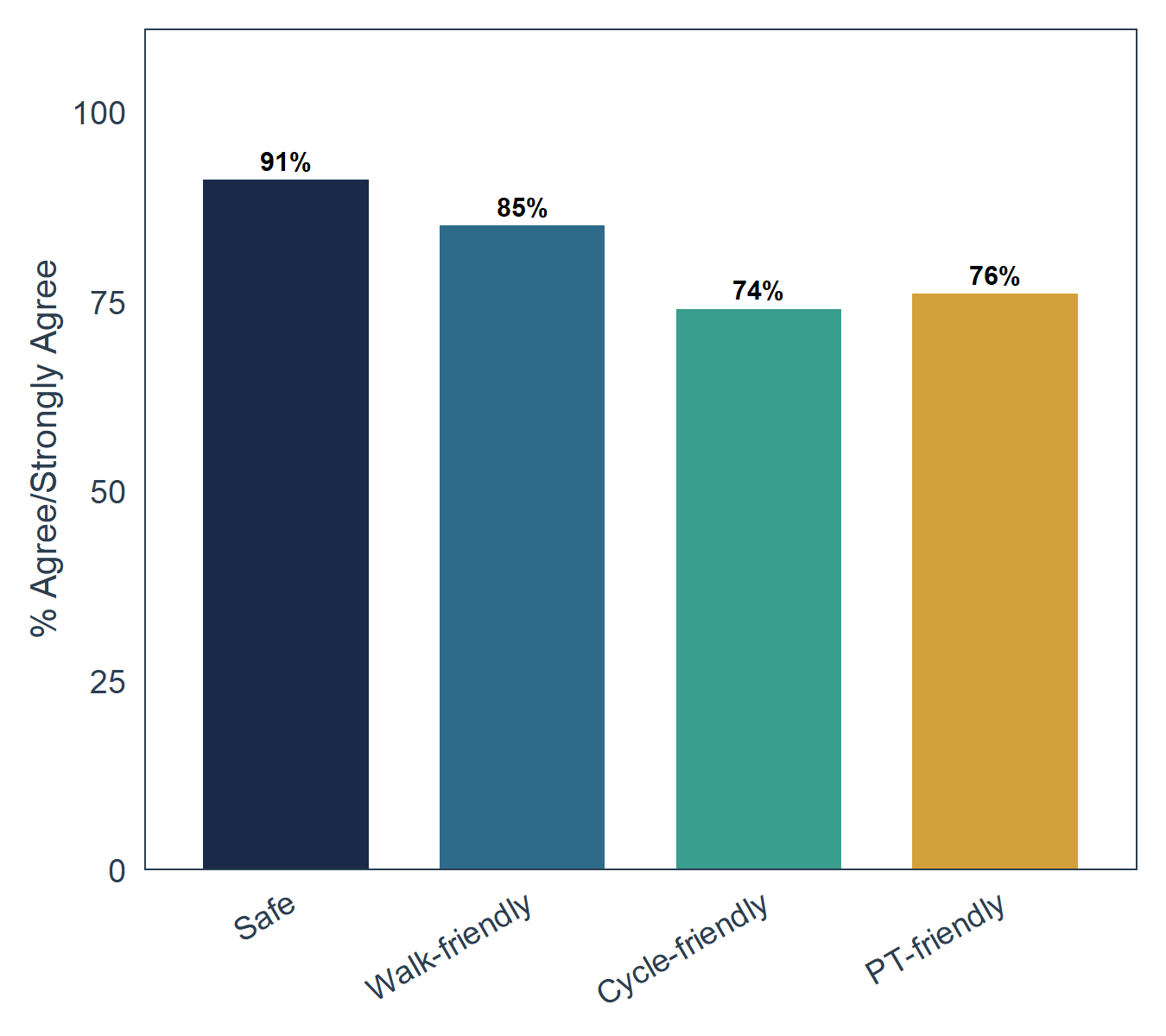}\caption{Residential perceptions}\label{fig:perceptions}\end{subfigure}
\caption{Sociodemographic profile of study participants by immigrant status.}
\label{fig:socio}
\end{figure}

Figure~\ref{fig:socio} summarizes the sociodemographic profile of the sample by immigrant status. The sample shows an overall higher representation of male participants (70\% overall), though the gender imbalance is less pronounced among immigrants (50\% male) than among Canadian-born participants (79\% male). Such imbalances are consistent with well-documented patterns in travel behaviour research, where both survey-based and technology-enabled data collection approaches are subject to demographic biases. Previous studies have shown that mobility datasets may systematically overrepresent certain groups, while travel surveys, particularly those employing smartphone-based or longitudinal designs, exhibit varying participation and engagement patterns across genders \citep{wesolowski2014quantifying, greaves2025stays, azoulay2024towards}. Figure~\ref{fig:socio}\subref{fig:age} shows the age distribution where Canadian-born participants cluster in the 45--54 bracket (49\%), whereas immigrants are more evenly distributed, with the largest share in the 35--44 bracket (31\%) followed by 25--34 (28\%) and 18--24 (25\%), which is consistent with Canada's immigration intake which favours working-age adults through economic pathways. Employment patterns reflect this age differential as shown in Figure~\ref{fig:socio}\subref{fig:employment}, where 72\% of Canadian-born participants work full-time compared to 38\% of immigrants, while 47\% of immigrants are students compared to only 13\% of Canadian-born. Marital status is balanced across groups (approximately 65\% married in both). Car ownership is high overall (83\% of households own at least one vehicle), though immigrants are substantially more likely to be car-free (41\% own zero cars, compared to 6\% of Canadian-born), as shown in Figure~\ref{fig:socio}\subref{fig:carown}. Both groups show a uniform positive residential perceptions shown in Figure~\ref{fig:socio}\subref{fig:perceptions}, with over 74\% of participants rating their neighbourhood favourably on safety, walkability, cycling infrastructure, and public transit access, which reflects the urban residential locations concentrated in central Toronto and Montreal.
 
\begin{figure}[htbp]
\centering
\begin{subfigure}[b]{0.32\textwidth}\centering\includegraphics[width=\textwidth]{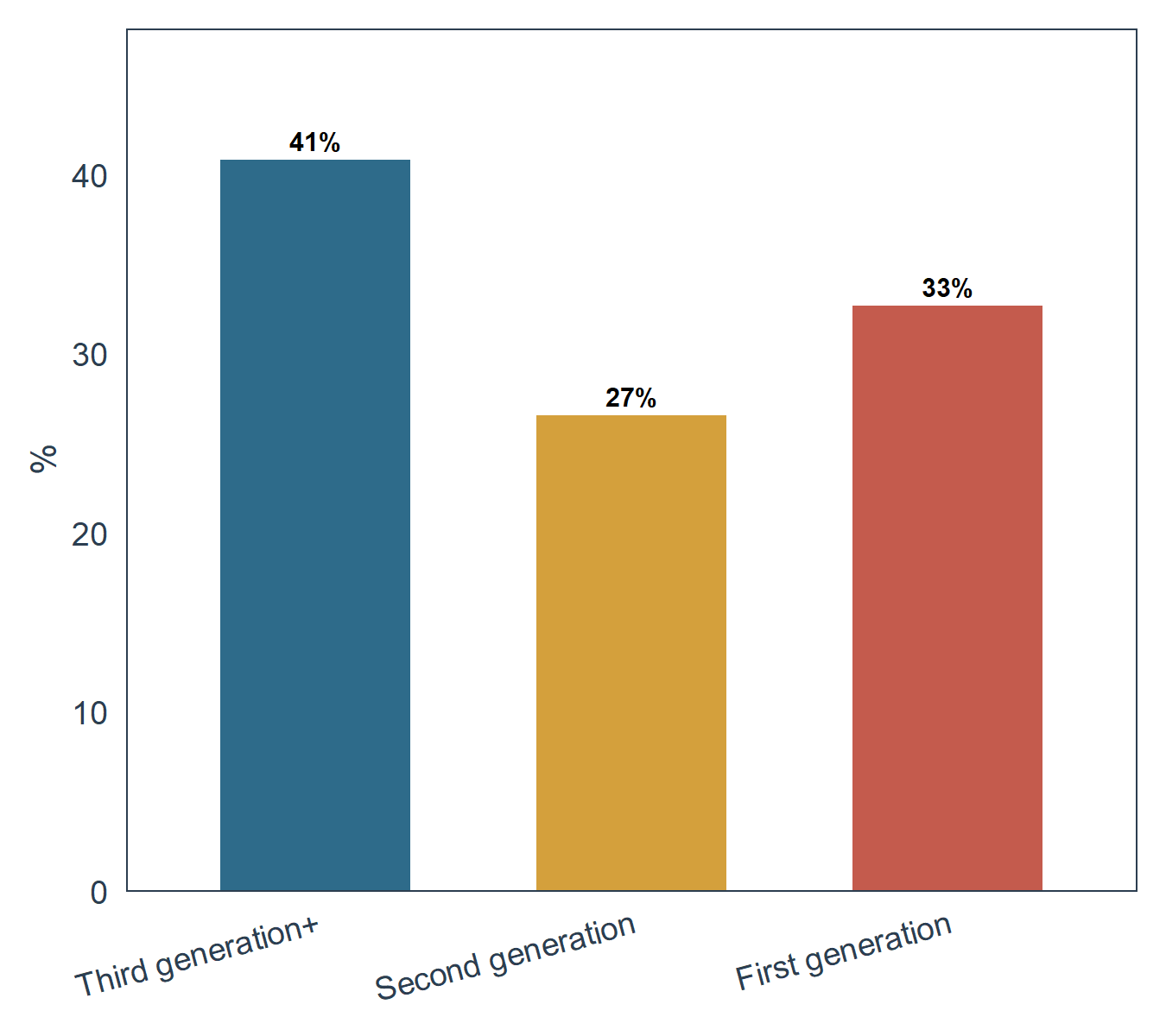}\caption{immigrant status}\label{fig:migstatus}\end{subfigure}\hfill
\begin{subfigure}[b]{0.32\textwidth}\centering\includegraphics[width=\textwidth]{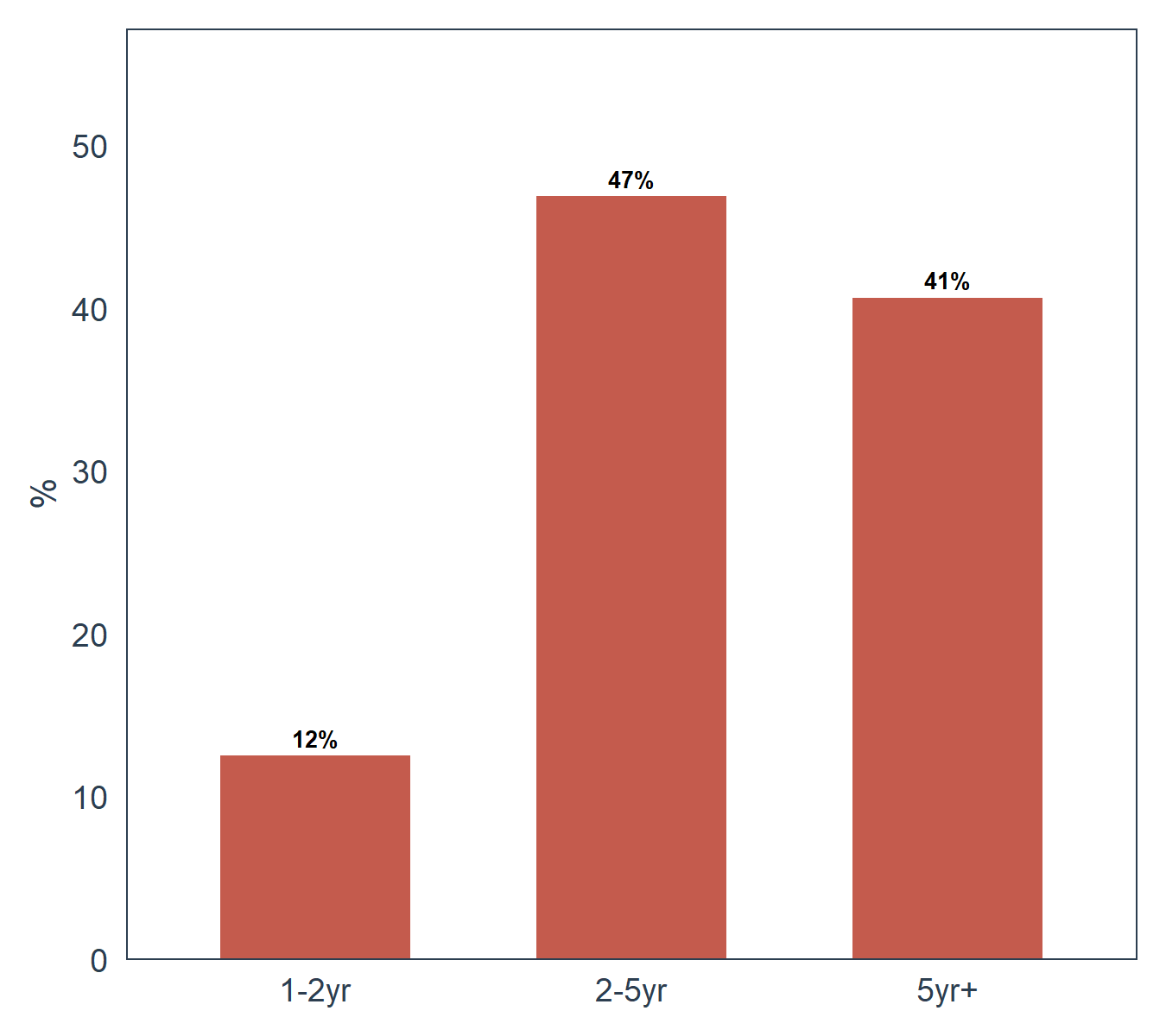}\caption{Time in Canada}\label{fig:timeincanada}\end{subfigure}\hfill
\begin{subfigure}[b]{0.32\textwidth}\centering\includegraphics[width=\textwidth]{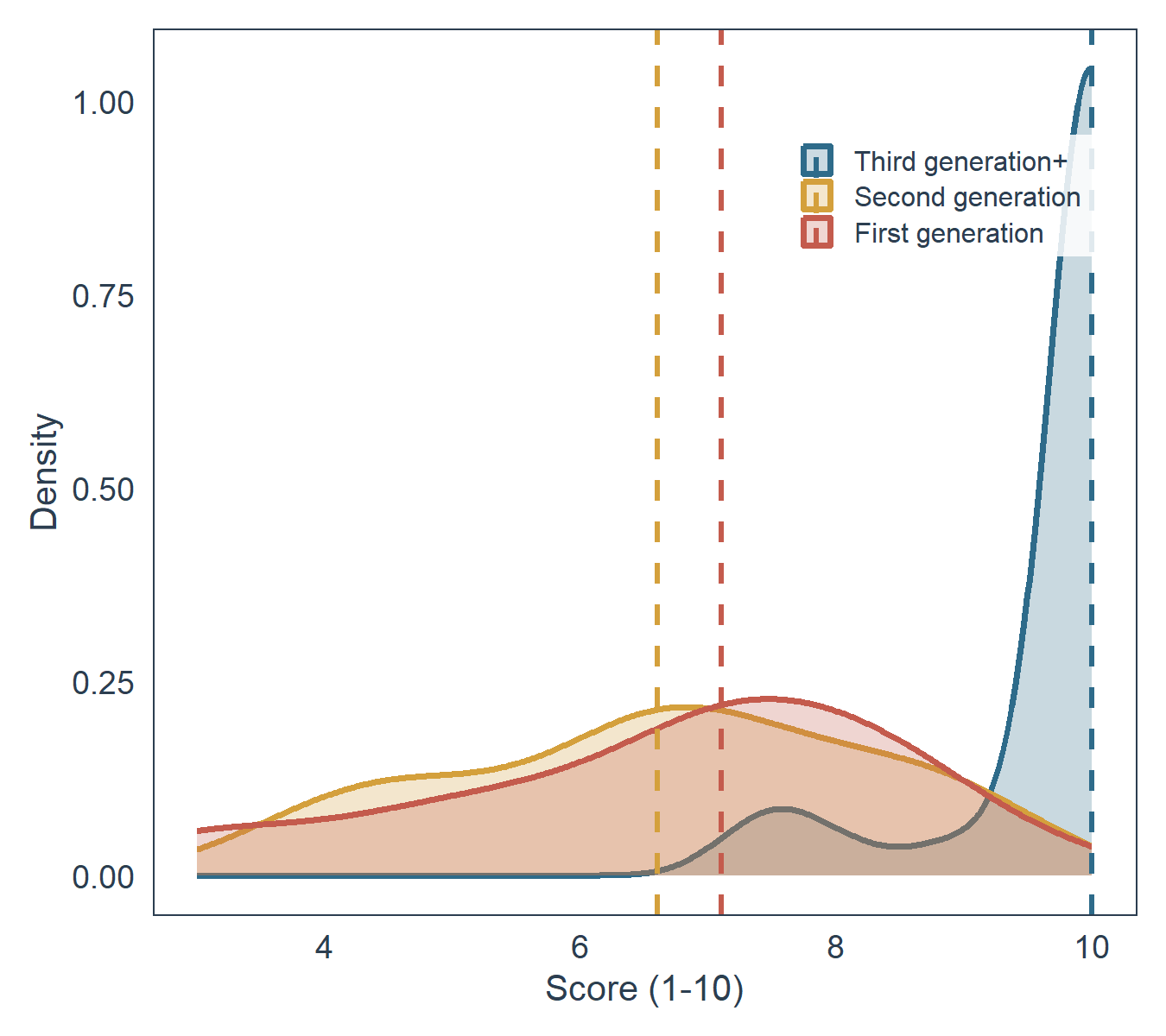}\caption{Integration density}\label{fig:integdensity}\end{subfigure}\\[6pt]
\begin{subfigure}[b]{0.32\textwidth}\centering\includegraphics[width=\textwidth]{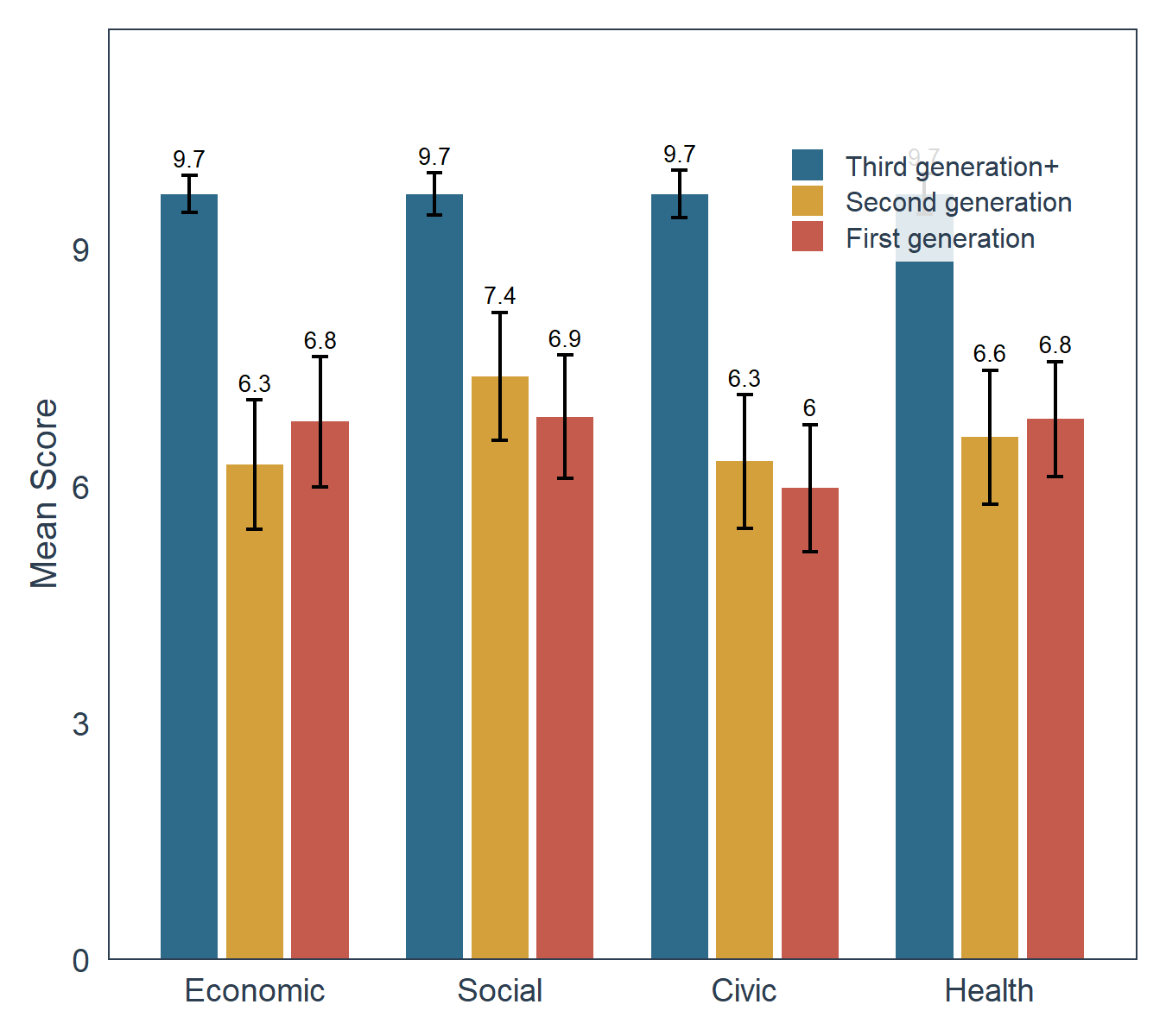}\caption{Sub-dimensions}\label{fig:subdimensions}\end{subfigure}\hfill
\begin{subfigure}[b]{0.32\textwidth}\centering\includegraphics[width=\textwidth]{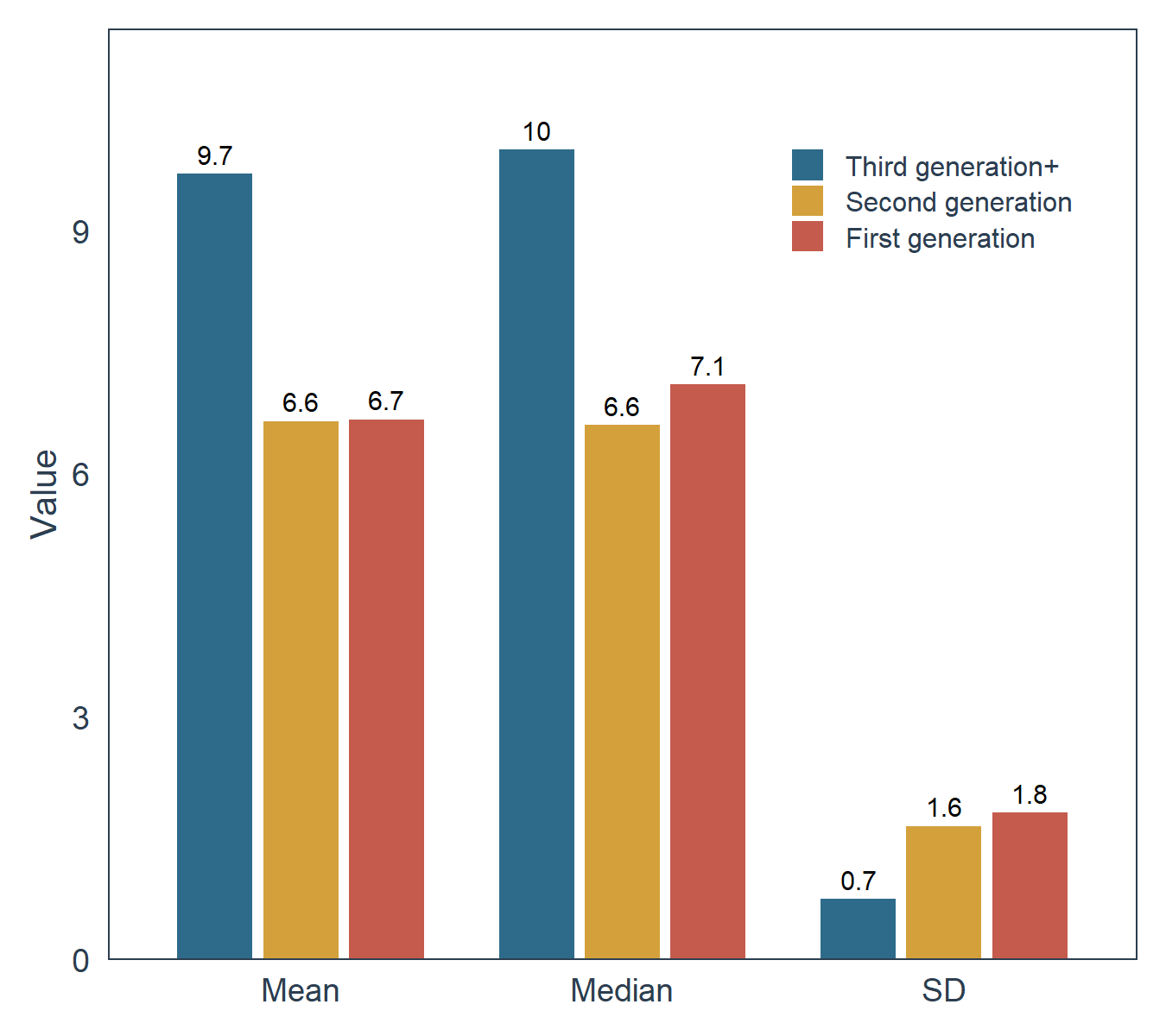}\caption{Index summary}\label{fig:integsummary}\end{subfigure}\hfill
\begin{subfigure}[b]{0.32\textwidth}\centering\includegraphics[width=\textwidth]{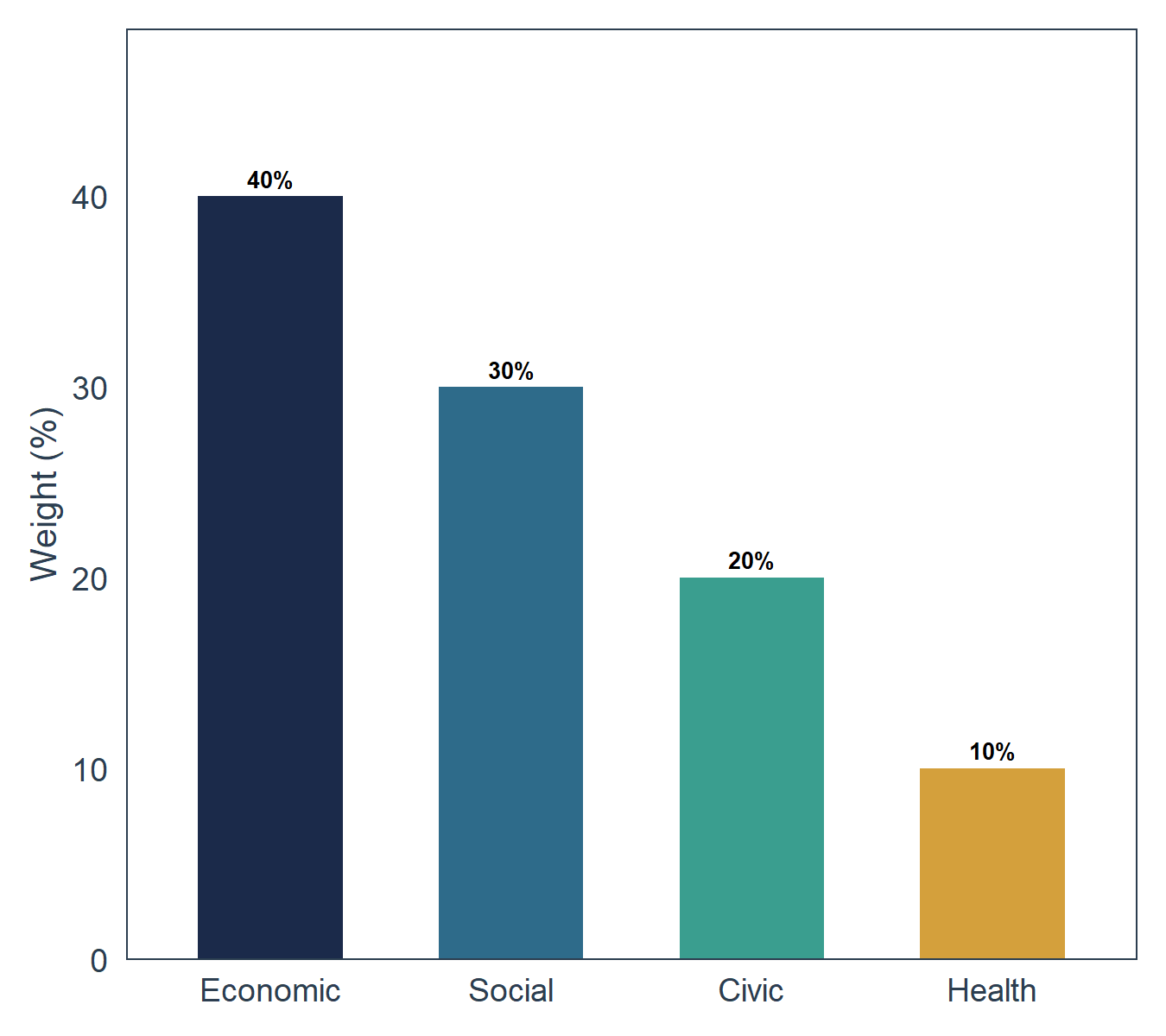}\caption{Dimension weights}\label{fig:weights}\end{subfigure}
\caption{immigrant status and integration indicators: \subref{fig:migstatus} immigrant status, \subref{fig:timeincanada} time in Canada (immigrants only), \subref{fig:integdensity} integration index density by group with median lines, \subref{fig:subdimensions} sub-dimension means with 95\% CIs, \subref{fig:integsummary} summary statistics, and \subref{fig:weights} dimension weights.}
\label{fig:integration}
\end{figure}

Of the 100 participants, 33\% are first-generation immigrants (born outside Canada), 27\% are second-generation (born in Canada with at least one foreign-born parent), and 41\% are third-generation or higher (both parents also born in Canada), as presented in Figure~\ref{fig:integration}\subref{fig:migstatus}. The proportion of foreign-born participants falls within the range observed across the study regions, where first-generation immigrants constitute approximately 48\% of the population in the Greater Toronto Area and 24.7\% in the Greater Montr\'{e}al Area \citep{statcan2021_census_profile}. Among the sampled first-generation immigrants, Figure~\ref{fig:integration}\subref{fig:timeincanada} shows the majority have resided in Canada for more than two years: 47\% for 2--5 years and 41\% for more than five years, with only 12\% in the 1--2 year category. The integration index, constructed as described in Section~\ref{sec:datacollection} on a 1--10 scale, reveals a clear gradient between groups presented in Figure~\ref{fig:integration}\subref{fig:integdensity}. Canadian-born participants report a mean index of 8.5 (median 9.7), while first generation immigrants report a mean of 6.7 (median 7.1), a gap of 1.8 points, or approximately 0.9 standard deviations on the centred index used in estimation. At the sub-dimension level (Figure~\ref{fig:integration}\subref{fig:subdimensions}), first generation immigrants score consistently lower across all four domains, with the largest gap in civic integration (6.0 vs.\ 8.4 for Canadian-born) and the smallest in health (6.8 vs.\ 8.5). The lower civic integration score likely reflects barriers to political participation among non-citizens.
 
The application of the integration index to both immigrant and Canadian-born participants is a deliberate methodological choice. The index measures the degree of socioeconomic and civic embeddedness in mainstream Canadian urban life, a dimension that varies meaningfully within both populations. For immigrants, lower scores reflect the ongoing settlement process such as language barriers, context experience, education credentials, unfamiliar institutions, and developing social networks. For Canadian-born individuals, lower scores may reflect socioeconomic marginalization, social isolation, or limited civic engagement. In the discrete choice models, two variables operate separately to capture different effects. The binary immigrant (first generation) vs Canadian-born dummy variable captures status-specific effects, while INTEG\_C captures the continuous embeddedness gradient for first generation immigrants and Canadian-born (second and third + generations) that moderates mode utility for all individuals.
 
\begin{figure}[htbp]
\centering
\begin{subfigure}[b]{0.32\textwidth}\centering\includegraphics[width=\textwidth]{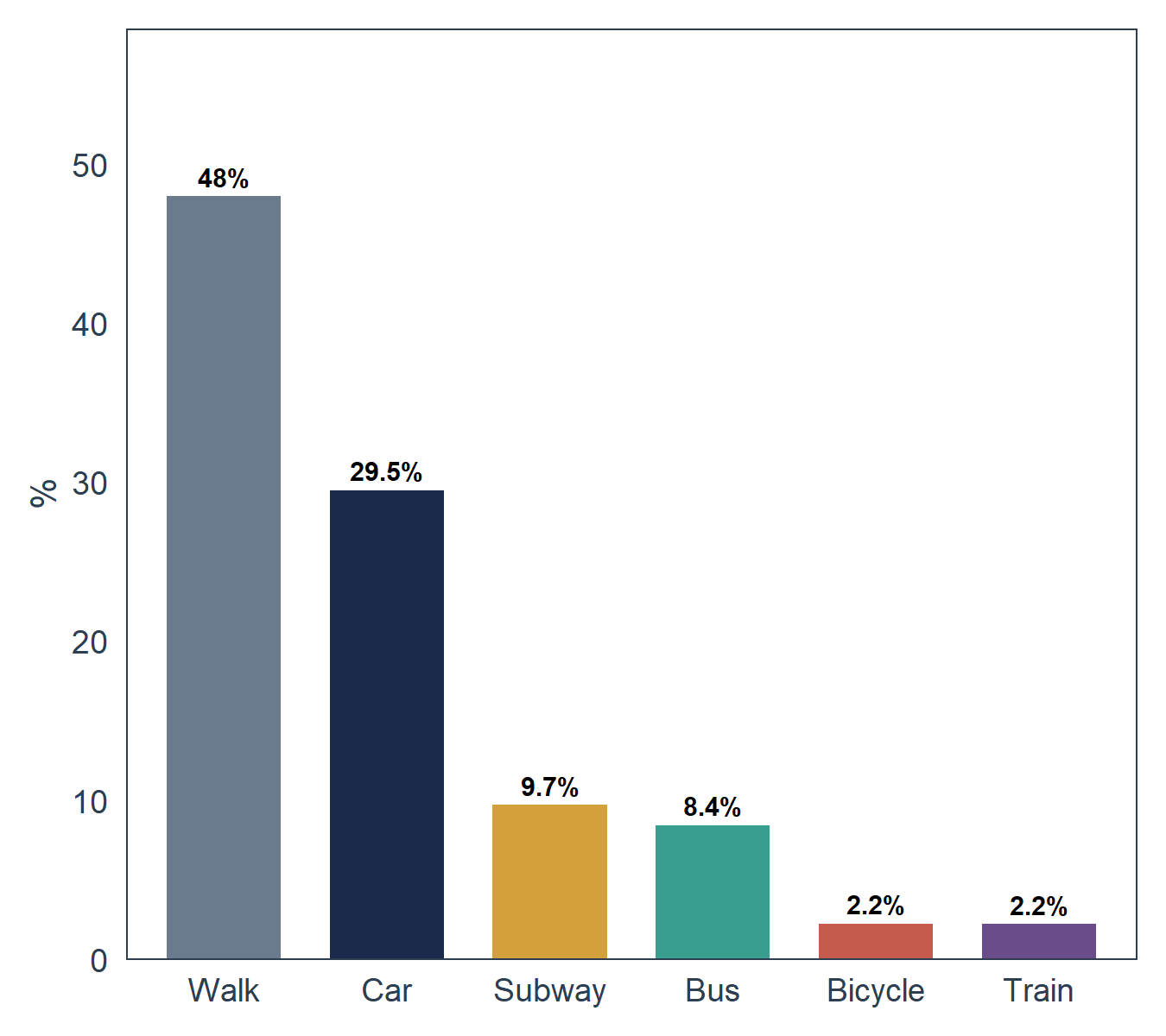}\caption{Share by trip count}\label{fig:modesharecount}\end{subfigure}\hfill
\begin{subfigure}[b]{0.32\textwidth}\centering\includegraphics[width=\textwidth]{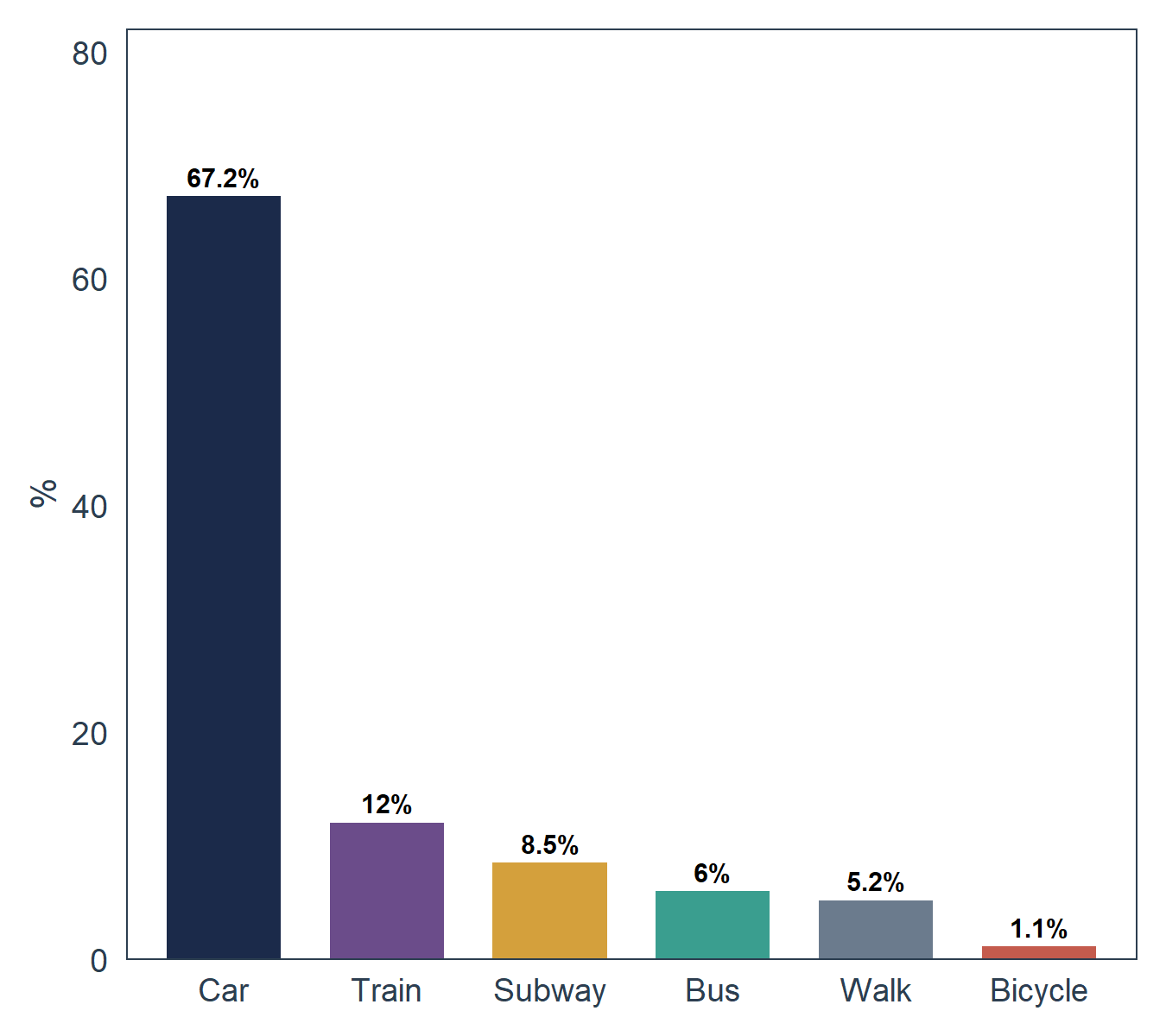}\caption{Share by distance}\label{fig:modesharedist}\end{subfigure}\hfill
\begin{subfigure}[b]{0.32\textwidth}\centering\includegraphics[width=\textwidth]{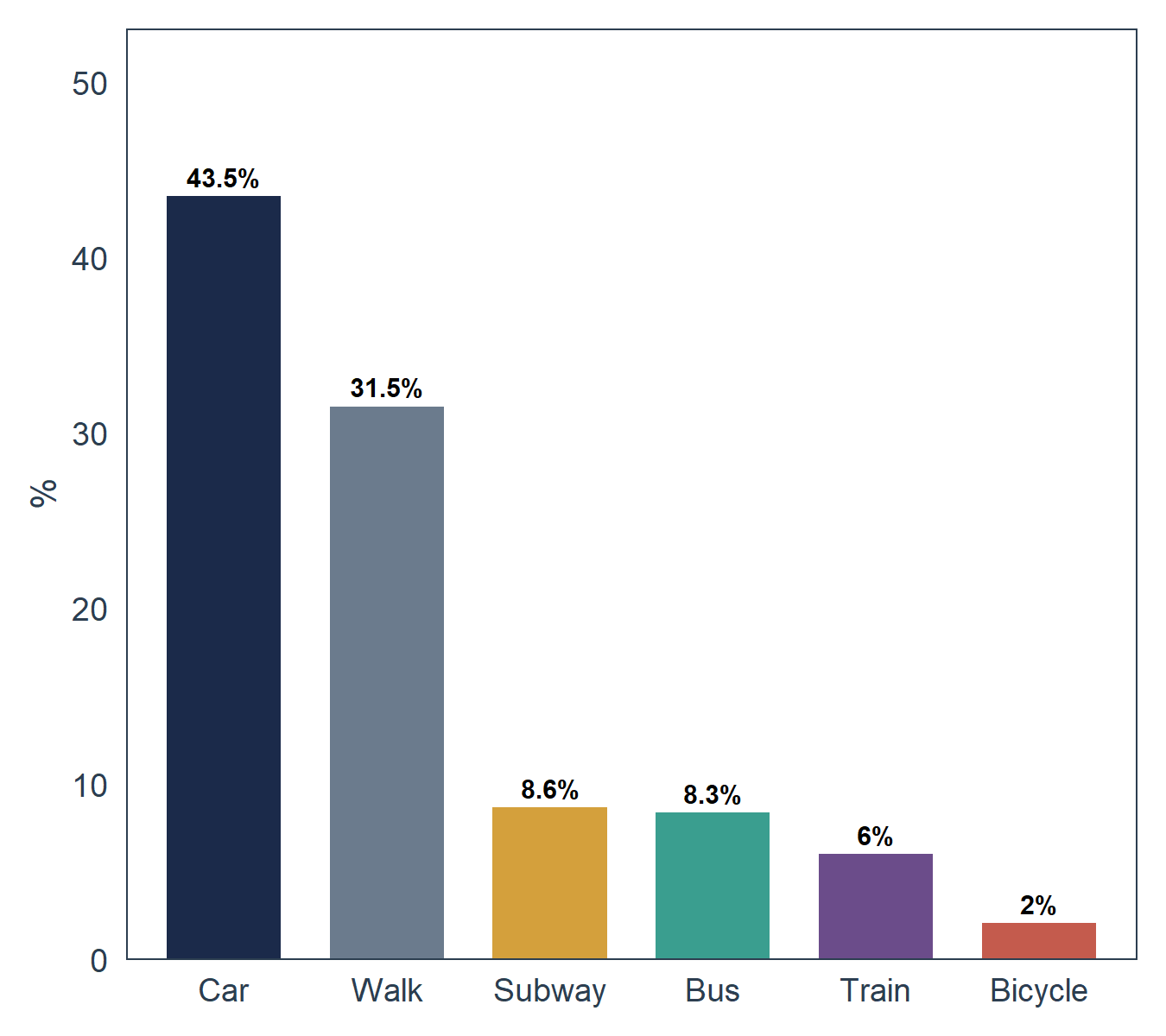}\caption{Share by travel time}\label{fig:modesharetime}\end{subfigure}\\[6pt]
\begin{subfigure}[b]{0.32\textwidth}\centering\includegraphics[width=\textwidth]{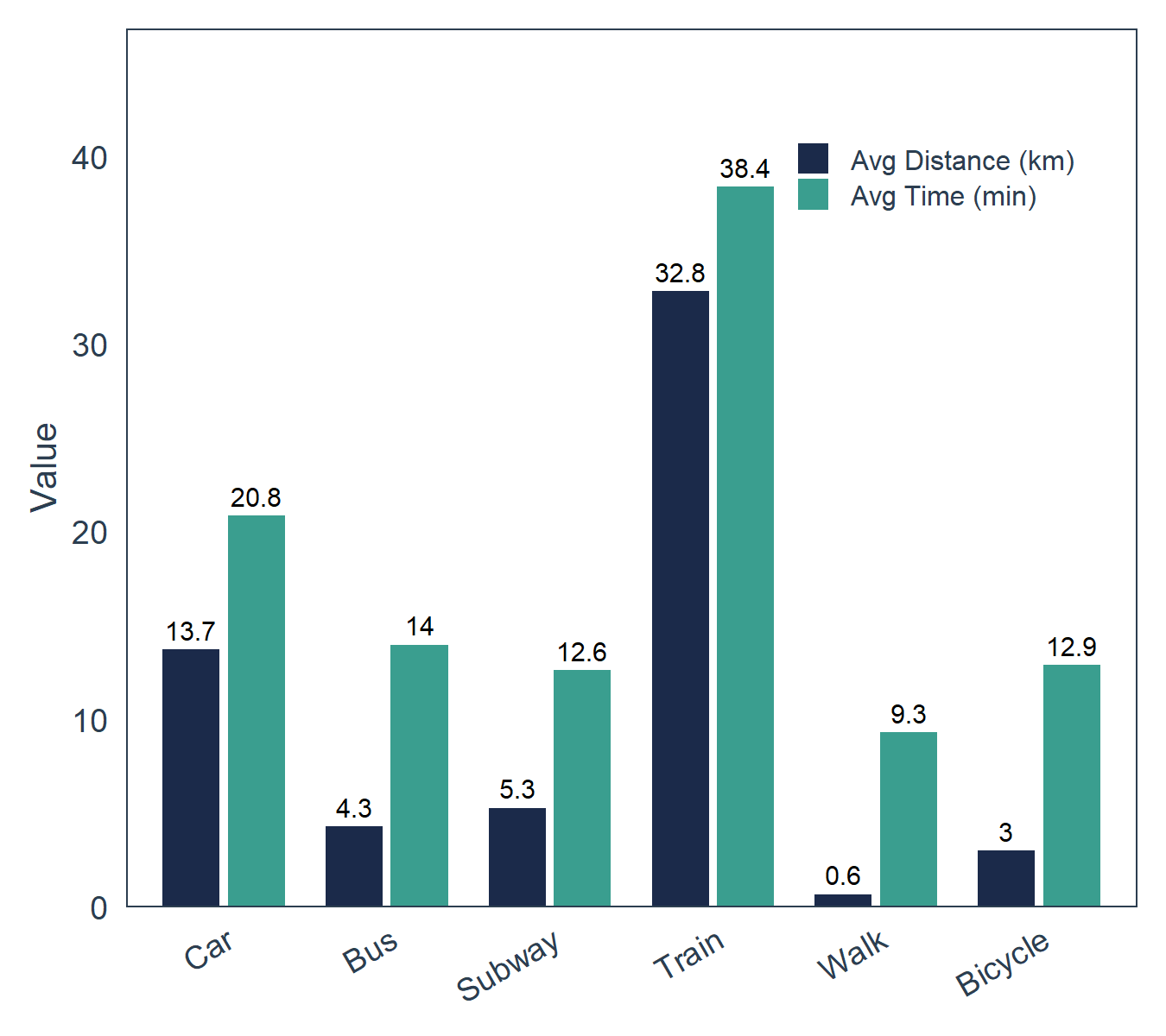}\caption{Avg distance and time}\label{fig:avgdisttime}\end{subfigure}\hfill
\begin{subfigure}[b]{0.32\textwidth}\centering\includegraphics[width=\textwidth]{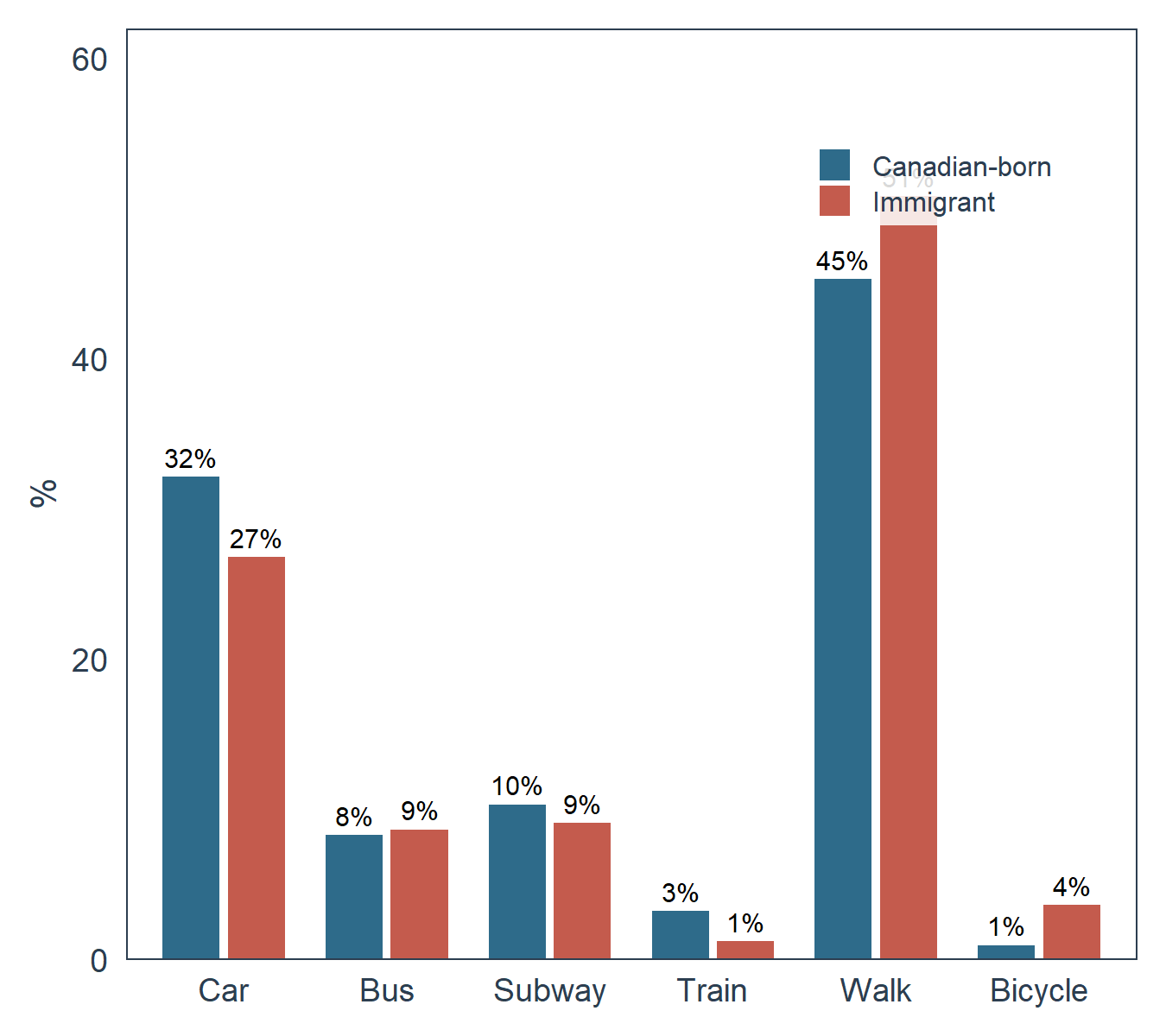}\caption{Share by immigrant status}\label{fig:modebymig}\end{subfigure}\hfill
\begin{subfigure}[b]{0.32\textwidth}\centering\includegraphics[width=\textwidth]{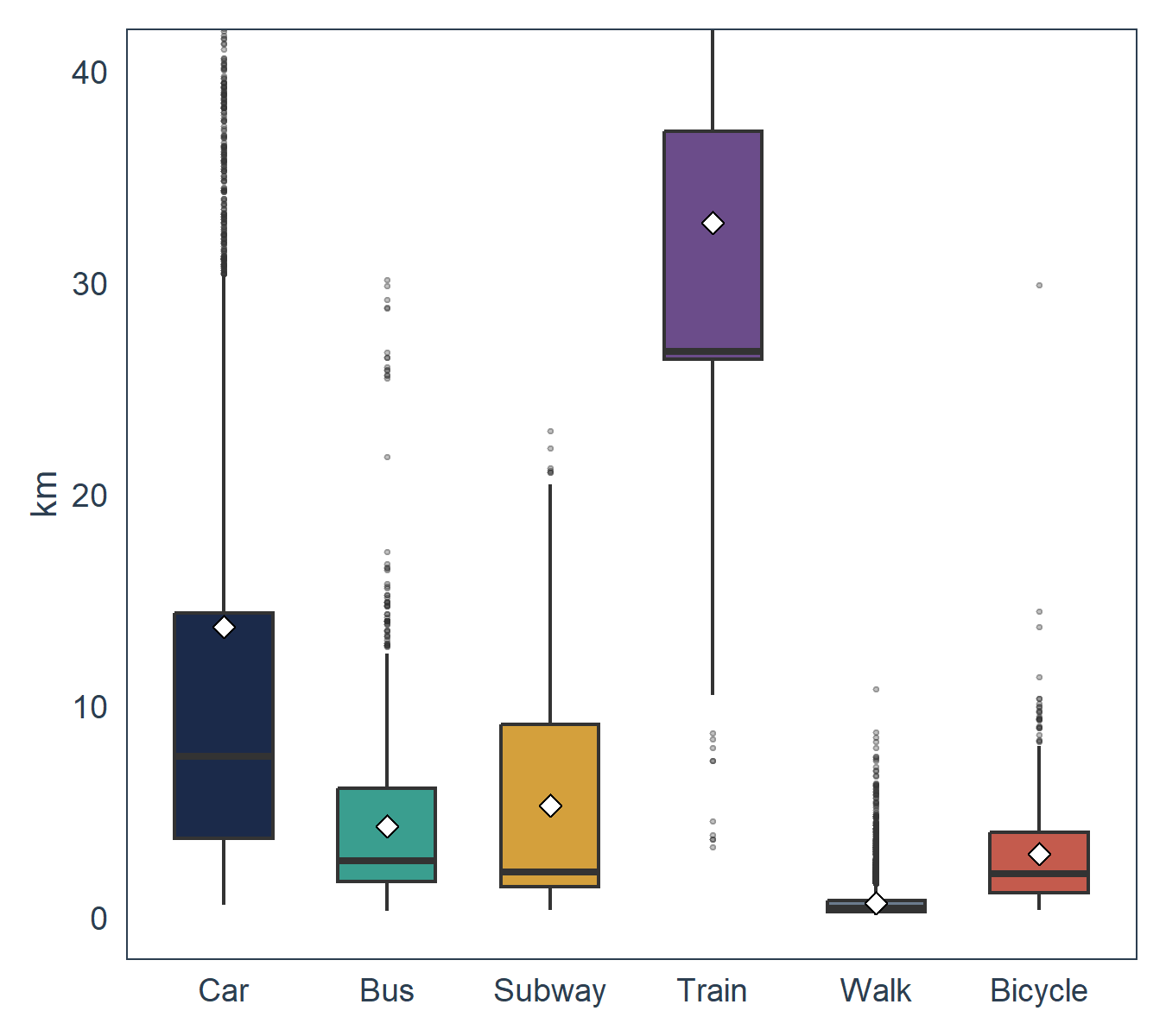}\caption{Distance distribution}\label{fig:distdist}\end{subfigure}
\caption{Recorded trip diary summary: \subref{fig:modesharecount} mode share by trip count, \subref{fig:modesharedist} by distance, \subref{fig:modesharetime} by travel time, \subref{fig:avgdisttime} average trip distance and duration by mode, \subref{fig:modebymig} mode share by immigrant status, and \subref{fig:distdist} trip distance distribution (box = IQR; line = median; diamond = mean).}
\label{fig:trips}
\end{figure}

Figure~\ref{fig:trips}\subref{fig:modesharecount}--\subref{fig:modesharetime} present mode share by trip count, total distance, and total travel time. Walking dominates trip count (49.7\%), followed by car (28.0\%), subway (9.6\%), bus (8.2\%), bicycle (2.3\%), and train (2.1\%). The high walk share is a distinctive feature of multi-day GPS panel data as continuous recording captures routine short trips that participants typically omit from single-day travel diaries. When measured by distance, car dominates (66.1\%), reflecting longer motorized trip distances, followed by train (12.1\%) and subway (8.8\%). Average trip characteristics vary substantially by mode as shown in Table~\ref{tab:tripattr}.  Walking trips average 0.6~km and 9.3 minutes, while train trips are the longest at 32.6~km and 38.2 minutes. Public transit fares are approximately flat at \$3.50 CAD, which is a result of the flat-fare structure of the GTA and GMA transit systems, while average car cost (fuel plus parking) is \$13.90 CAD.
 
\begin{table}[htbp]
\centering
\caption{Trip attributes by mode}
\label{tab:tripattr}
\small
\begin{tabular}{lrrrrr}
\toprule
\textbf{Mode} & \textbf{Trips} & \textbf{Dist.\ mean (km)} & \textbf{Dist.\ med.\ (km)} & \textbf{Time mean (min)} & \textbf{Time med.\ (min)} \\
\midrule
Car     & 4{,}979 & 13.5 & 7.5 & 20.7 & 16.0 \\
Bus     & 1{,}461 &  4.2 & 2.6 & 13.9 & 11.2 \\
Subway  & 1{,}713 &  5.2 & 2.0 & 12.6 &  6.9 \\
Train   &    379  & 32.6 & 26.7 & 38.2 & 34.7 \\
Walk    & 8{,}841 &  0.6 & 0.4 &  9.3 &  6.0 \\
Bicycle &    411  &  3.0 & 2.0 & 12.9 & 10.0 \\
\bottomrule
\end{tabular}
\end{table}
 
The disaggregation of the travel diaries by immigrant status reveals patterns consistent with the integration hypothesis as presented in Figure~\ref{fig:trips}\subref{fig:modebymig}. Canadian-born participants exhibit higher car mode share (30\% vs.\ 26\% for immigrants) and higher train usage (3\% vs.\ 1\%), while immigrants show higher walking share (53\% vs.\ 47\%) and notably higher cycling (4\% vs.\ 1\%). Bus and subway shares are similar across groups. These raw differences suggest that immigrants rely more on non-motorized modes, though they confound factors such as car access, residential location, and trip distance that are disentangled in the econometric models.
 
\begin{figure}[htbp]
\centering
\begin{subfigure}[b]{0.48\textwidth}\centering\includegraphics[width=\textwidth]{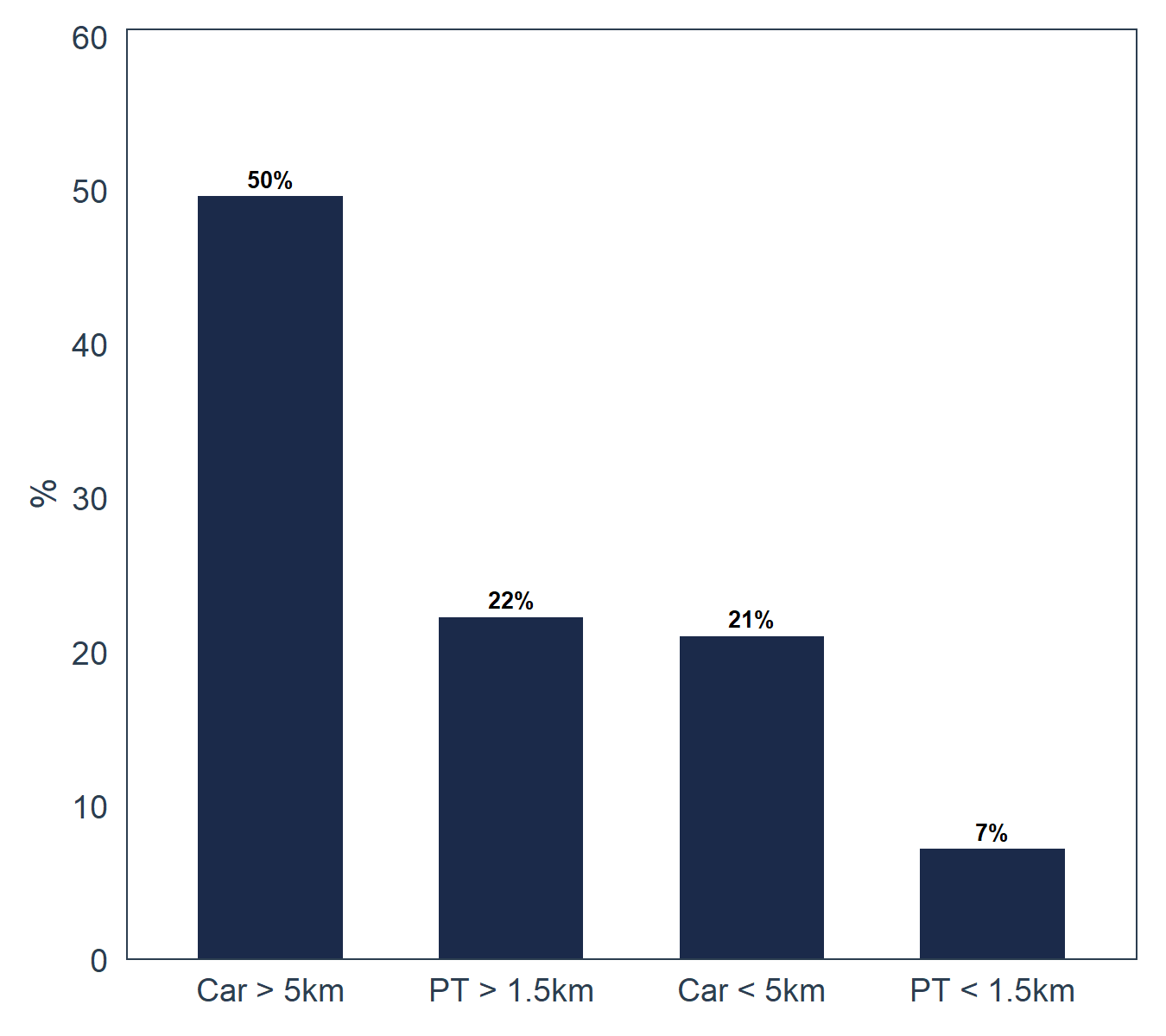}\caption{SP scenario types}\label{fig:sptypes}\end{subfigure}\hfill
\begin{subfigure}[b]{0.48\textwidth}\centering\includegraphics[width=\textwidth]{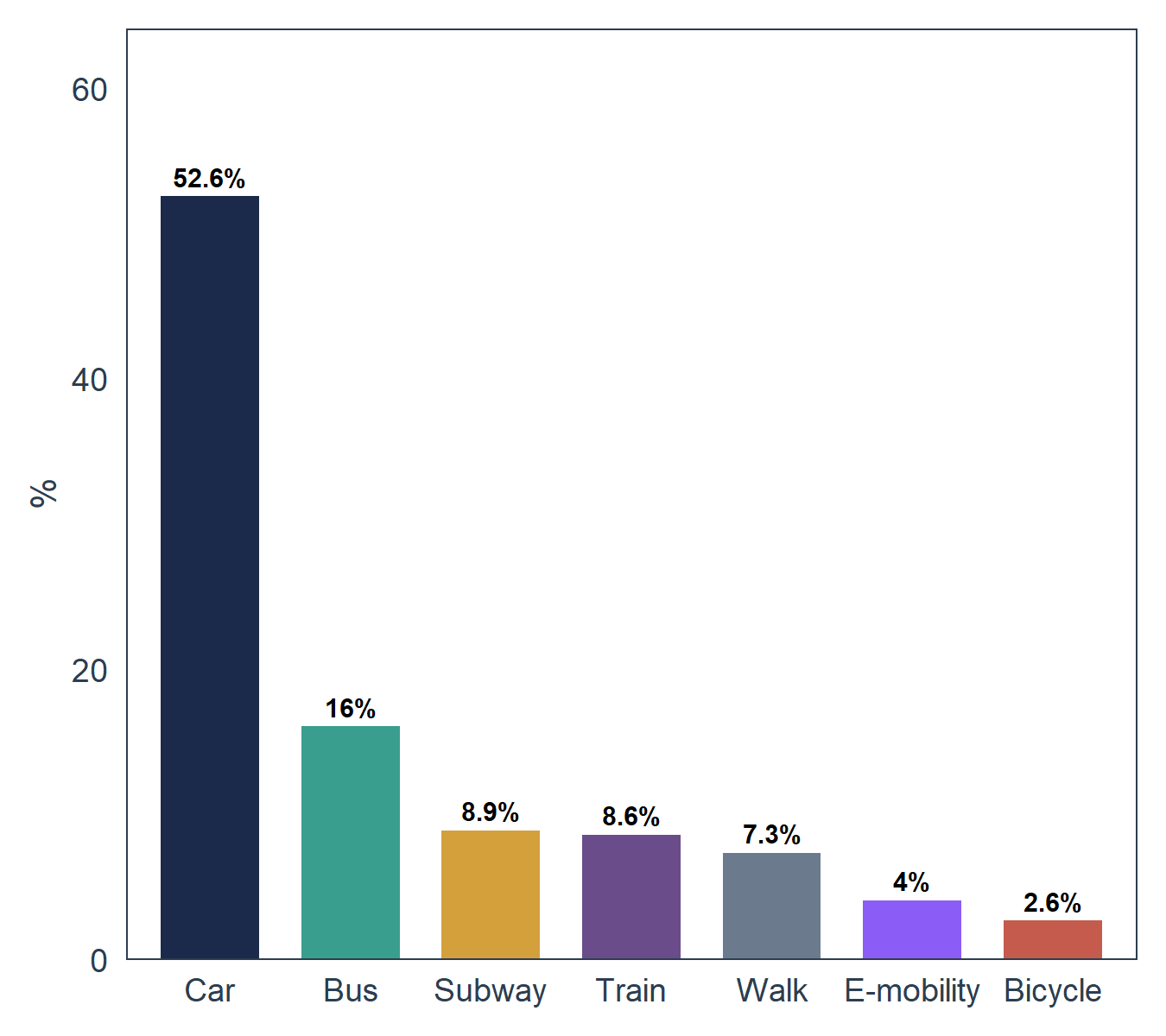}\caption{SP mode chosen}\label{fig:spmodechosen}\end{subfigure}\\[6pt]
\begin{subfigure}[b]{0.48\textwidth}\centering\includegraphics[width=\textwidth]{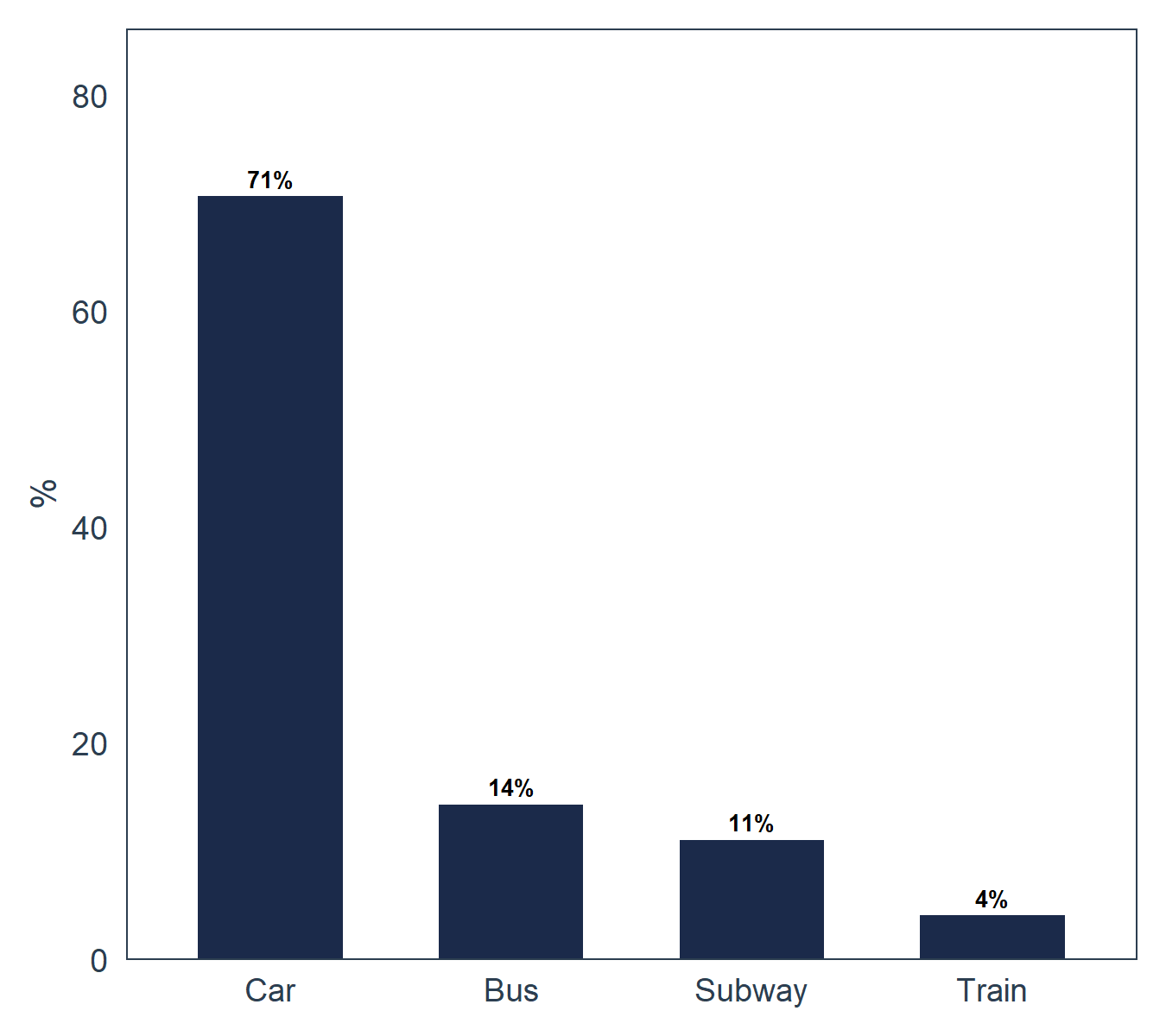}\caption{Original RP mode}\label{fig:originalrpmode}\end{subfigure}\hfill
\begin{subfigure}[b]{0.48\textwidth}\centering\includegraphics[width=\textwidth]{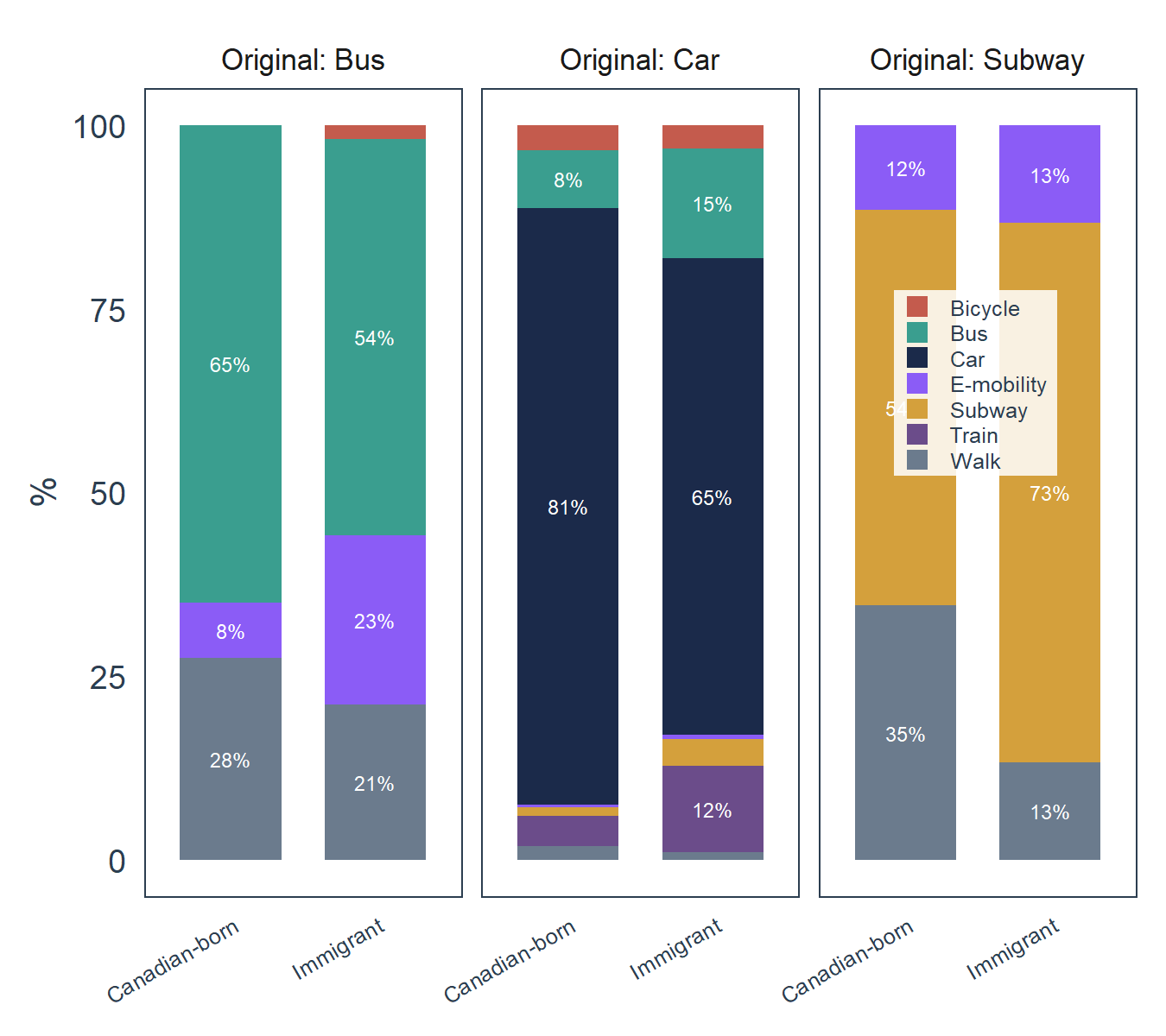}\caption{Alternative by immigrant status}\label{fig:chosenaltbygroup}\end{subfigure}
\caption{Stated preference scenarios and behavioural shift: \subref{fig:sptypes} distribution of SP scenario types, \subref{fig:spmodechosen} overall SP mode chosen, \subref{fig:originalrpmode} original RP mode of SP-triggered trips, and \subref{fig:chosenaltbygroup} chosen alternative per detected mode by immigrant status.}
\label{fig:sp}
\end{figure}

The SP scenarios were triggered dynamically within the app based on participants' actual observed trips, presenting hypothetical mode-switching alternatives during daily travel. Half of all scenarios (50\%) were triggered on car trips exceeding 5~km, with smaller shares from public transit trips above 1.5~km (22\%), car trips below 5~km (21\%), and public transit trips below 1.5~km (7\%). In the stated choices presented in Figure~\ref{fig:sp}\subref{fig:spmodechosen}, 52.6\% of respondents chose to retain their current mode, 16.0\% selected bus, 8.9\% subway, 8.6\% train, 7.3\% walking, 4.0\% e-mobility options, and 2.6\% bicycle as alternatives to their observed modes.
 
Figure~\ref{fig:sp}\subref{fig:chosenaltbygroup} decomposes the chosen alternative by original detected mode and immigrant status. When the original mode is \emph{bus}, Canadian-born participants are more likely to retain bus (65\%) compared to immigrants (54\%), while immigrants show greater willingness to shift to e-mobility options (23\% vs.\ 8\%) and less willingness to substitute bus trips with walking (28\% vs.\ 21\%). For \emph{car} trips, Canadian-born participants overwhelmingly retain car (81\%), while immigrants show modestly more openness to alternatives. The starkest contrast emerges for \emph{subway} trips: immigrants strongly retain subway (73\%) whereas Canadian-born participants show a broader distribution across alternatives (subway retention 54\%, walk 35\%).
 
\subsection{Model Estimation and Validation}\label{sec:estimation}

Table~\ref{tab:modelfit} summarizes the estimation results and goodness-of-fit statistics for all four specifications, estimated on the full sample. All models exhibit substantial improvement over the null model, with adjusted $\rho^2$ values ranging from 0.455 to 0.532. The MXL models (M2, M4) achieve higher adjusted $\rho^2$ than their MNL counterparts despite having fewer parameters, which indicates that the inclusion of random parameters improves model fit by capturing unobserved preference heterogeneity and enhances the model’s ability to represent behavioural variation. Tables~\ref{tab:params_mnl} and~\ref{tab:params_mxl} report the full parameter estimates, with MNL models (M1 and M3) presented side-by-side to highlight the effect of adding SP data, and MXL models (M2 and M4) similarly paired.
 
\begin{table}[htbp]
\centering
\caption{Model fit comparison}
\label{tab:modelfit}
\footnotesize
\begin{tabular}{lcccc}
\toprule
& \textbf{Model 1} & \textbf{Model 2} & \textbf{Model 3} & \textbf{Model 4} \\
& MNL RP & MXL RP & MNL RP-SP & MXL RP-SP \\
\midrule
Parameters            & 21        & 17        & 24         & 19 \\
Observations          & 14{,}502  & 10{,}241  & 15{,}124   & 9{,}455 \\
$\mathcal{LL}(\text{final})$ & $-$6{,}842.8 & $-$4{,}257.6 & $-$7{,}487.5 & $-$4{,}254.9 \\
Adj.\ $\rho^2(0)$    & 0.473     & 0.532     & 0.455      & 0.504 \\
AIC                   & 13{,}728  & 8{,}549   & 15{,}023   & 8{,}548 \\
Random parameters     & ---       & 2         & ---        & 2 \\
$\mu_{\text{SP}}$     & ---       & ---       & 0.298      & 0.281 \\
E-mobility            & ---       & ---       & $\checkmark$ & $\checkmark$ \\[3pt]
\multicolumn{5}{l}{\textit{Five-fold observation-level cross-validation}} \\
CV accuracy (mean $\pm$ SD) & 82.0 $\pm$ 0.7\% & 79.9 $\pm$ 0.7\% & 81.8 $\pm$ 0.8\% & 79.6 $\pm$ 0.7\% \\
\bottomrule
\multicolumn{5}{p{12.5cm}}{\footnotesize MXL observation counts reflect stratified trip caps (300 per person for M2,and M4) applied to prevent panel-product underflow. All SP observations retained in M3 and M4.}
\end{tabular}
\end{table}
 
\begin{table}[htbp]
\centering
\caption{MNL parameter estimates: Model~1 (RP) and Model~3 (Joint RP-SP)}
\label{tab:params_mnl}
\footnotesize
\begin{tabular}{l rr rr}
\toprule
& \multicolumn{2}{c}{\textbf{Model 1: MNL RP}} & \multicolumn{2}{c}{\textbf{Model 3: MNL RP-SP}} \\
\cmidrule(lr){2-3} \cmidrule(lr){4-5}
\textbf{Parameter} & Est. & Rob.\ $t$ & Est. & Rob.\ $t$ \\
\midrule
\multicolumn{5}{l}{\textit{Scale parameter}} \\
$\mu_{\text{SP}}$ & --- & --- & 0.298 & 5.51$^{***}$ \\[3pt]
\multicolumn{5}{l}{\textit{Alternative-specific constants}} \\
$\alpha_{\text{bus}}$    & $-$0.271 & $-$3.31$^{***}$ & $-$0.310 & $-$3.75$^{***}$ \\
$\alpha_{\text{sub}}$    &    2.211 &   18.09$^{***}$  &    1.873 &    8.94$^{***}$ \\
$\alpha_{\text{walk}}$   &    0.965 &   10.18$^{***}$  &    0.888 &    9.19$^{***}$ \\
$\alpha_{\text{bike}}$   & $-$2.234 & $-$14.32$^{***}$ & $-$2.314 & $-$14.67$^{***}$ \\
$\alpha_{\text{emob}}$   & --- & ---                    & $-$1.294 & $-$1.57 \\[3pt]
\multicolumn{5}{l}{\textit{Level-of-service}} \\
$\beta_{C}$ (Cost)       & $-$0.766 & $-$15.02$^{***}$ & $-$0.801 & $-$15.59$^{***}$ \\
$\beta_{T}$ (IVTT)       & $-$0.361 &  $-$9.94$^{***}$ & $-$0.347 &  $-$9.51$^{***}$ \\
$\beta_{A}$ (Walk time)  & $-$0.461 &  $-$9.33$^{***}$ & $-$0.455 &  $-$9.15$^{***}$ \\[3pt]
\multicolumn{5}{l}{\textit{Distance}} \\
$\beta_{D1}$ (Car)       &    0.356 &   20.55$^{***}$  &    0.353 &   20.62$^{***}$ \\
$\beta_{D2}$ (PT)        &    0.106 &    4.84$^{***}$  &    0.108 &    5.05$^{***}$ \\
$\beta_{D3}$ (Train)     &    0.311 &   18.44$^{***}$  &    0.302 &   18.05$^{***}$ \\
$\beta_{D4}$ (Active)    & $-$0.581 &  $-$8.98$^{***}$ & $-$0.575 &  $-$8.92$^{***}$ \\[3pt]
\multicolumn{5}{l}{\textit{Sociodemographic interactions}} \\
$\beta_{W}$ (Work/study $\to$ Car) & $-$0.291 & $-$4.22$^{***}$ & $-$0.252 & $-$3.68$^{***}$ \\
$\beta_{Ch}$ (Child $\to$ PT)      &    0.426 &    4.95$^{***}$  &    0.355 &    4.11$^{***}$ \\
$\beta_{M}$ (immigrant $\to$ Sub)   & $-$0.708 & $-$7.43$^{***}$ & $-$0.653 & $-$6.54$^{***}$ \\
$\beta_{F}$ (Full-time $\to$ Sub)  & $-$0.306 & $-$3.07$^{***}$ & $-$0.246 & $-$2.36$^{**}$ \\
$\beta_{St}$ (Student $\to$ Train) & $-$1.562 & $-$8.22$^{***}$ & $-$1.534 & $-$8.07$^{***}$ \\
$\beta_{Sw}$ (Student $\to$ Walk)  &    0.574 &    9.11$^{***}$  &    0.577 &    9.11$^{***}$ \\[3pt]
\multicolumn{5}{l}{\textit{Neighbourhood perceptions}} \\
$\beta_{S}$ (Safe $\to$ Sub)$^{\dagger}$ & --- & --- &    0.278 &    1.75$^{*}$ \\
$\beta_{Cy}$ (Cycle-fr.\ $\to$ Bike) & 0.704 & 4.96$^{***}$ &  0.720 &    5.03$^{***}$ \\[3pt]
\multicolumn{5}{l}{\textit{Integration and context}} \\
$\beta_{I1}$ (INTEG\_C $\to$ PT)  & $-$0.179 & $-$9.03$^{***}$ & $-$0.213 & $-$10.18$^{***}$ \\
$\beta_{I2}$ (INTEG\_C $\to$ Act) & $-$0.259 & $-$11.02$^{***}$ & $-$0.273 &  $-$11.56$^{***}$ \\
$\beta_{Sn}$ (Snow $\to$ Active)  & $-$0.622 & $-$2.46$^{**}$  & $-$0.660 &  $-$2.62$^{***}$ \\
\bottomrule
\multicolumn{5}{p{12.5cm}}{\footnotesize $^{***}p<0.01$, $^{**}p<0.05$, $^{*}p<0.10$. Robust $t$-statistics. Car and train ASCs normalized to 0. Reference categories: Canadian-born (MIG), part-time/other (FT), non-student (STU), no young children (CHILD), non-snow (SNOW).}
\end{tabular}
\end{table}
 
\begin{table}[htbp]
\centering
\caption{MXL parameter estimates: Model~2 (RP) and Model~4 (Joint RP-SP)}
\label{tab:params_mxl}
\footnotesize
\begin{tabular}{l rr rr}
\toprule
& \multicolumn{2}{c}{\textbf{Model 2: MXL RP}} & \multicolumn{2}{c}{\textbf{Model 4: MXL RP-SP}} \\
\cmidrule(lr){2-3} \cmidrule(lr){4-5}
\textbf{Parameter} & Est. & Rob.\ $t$ & Est. & Rob.\ $t$ \\
\midrule
\multicolumn{5}{l}{\textit{Scale parameter}} \\
$\mu_{\text{SP}}$ & --- & --- & 0.281 & 2.76$^{***}$ \\[3pt]
\multicolumn{5}{l}{\textit{Alternative-specific constants}} \\
$\alpha_{\text{bus}}$    & $-$0.396 & $-$1.16  & $-$0.519 & $-$1.50 \\
$\alpha_{\text{sub}}$    &    2.107 &    6.07$^{***}$ &    2.016 &    5.66$^{***}$ \\
$\alpha_{\text{walk}}$   &    1.088 &    3.00$^{***}$ &    0.834 &    2.27$^{**}$ \\
$\alpha_{\text{bike}}$   & $-$1.713 & $-$3.88$^{***}$ & $-$1.869 & $-$4.77$^{***}$ \\
$\alpha_{\text{emob}}$   & --- & ---                    & $-$0.810 & $-$0.35 \\[3pt]
\multicolumn{5}{l}{\textit{Random parameters}} \\
$\mu_T$ (IVTT mean)           & $-$0.848 & $-$8.57$^{***}$ & $-$0.798 & $-$7.21$^{***}$ \\
$\sigma_T$ (IVTT s.d.)        &    0.713 &   11.25$^{***}$  &    0.706 &    9.81$^{***}$ \\
$\delta_{\text{MIG}}$ (immigrant shift) & 0.556 & 5.25$^{***}$ & 0.525 & 3.58$^{***}$ \\
$\mu_C$ (Cost log-mean)       & $-$2.130 & $-$3.97$^{***}$ & $-$2.039 & $-$5.33$^{***}$ \\
$\sigma_C$ (Cost log-s.d.)    &    1.724 &    6.84$^{***}$  &    1.697 &    9.10$^{***}$ \\[3pt]
\multicolumn{5}{l}{\textit{Fixed parameters}} \\
$\beta_{A}$ (Walk time)       & $-$0.448 & $-$2.87$^{***}$ & $-$0.531 & $-$4.55$^{***}$ \\
$\beta_{D1}$ (Dist.\ Car)     &    0.394 &   10.46$^{***}$  &    0.382 &    8.95$^{***}$ \\
$\beta_{D2}$ (Dist.\ PT)      &    0.110 &    1.86$^{*}$    &    0.107 &    1.80$^{*}$ \\
$\beta_{D3}$ (Dist.\ Train)   &    0.276 &    7.04$^{***}$  &    0.251 &    5.60$^{***}$ \\
$\beta_{D4}$ (Dist.\ Active)  & $-$0.864 & $-$4.64$^{***}$ & $-$0.846 & $-$5.30$^{***}$ \\
$\beta_{F}$ (FT $\to$ Sub)    & $-$0.684 & $-$2.10$^{**}$  & $-$0.673 & $-$2.06$^{**}$ \\
$\beta_{Sw}$ (STU $\to$ Walk) &    0.559 &    2.10$^{**}$   &    0.656 &    2.20$^{**}$ \\
$\beta_{I1}$ (INTEG\_C $\to$ PT) & $-$0.095 & $-$0.55     & $-$0.106 & $-$0.73 \\
\bottomrule
\multicolumn{5}{p{12.5cm}}{\footnotesize 500 Halton draws. $^{***}p<0.01$, $^{**}p<0.05$, $^{*}p<0.10$. Robust $t$-statistics. Car and train ASCs normalized to 0. Reference categories: Canadian-born (MIG), part-time/other (FT), non-student (STU).}
\end{tabular}
\end{table}
 
All four specifications yield behaviourally consistent responses with respect to cost, travel time, and distance. The cost coefficient ranges from $-0.766$ (M1) to $-0.801$ (M3) in the MNL specifications, while in the MXL models cost is specified as a negative lognormal random parameter with log-mean $\mu_C \approx -2.1$. In-vehicle travel time is likewise negative and stable across specifications, with MNL estimates of $-0.361$ (M1) and $-0.347$ (M3), and corresponding MXL population means of $-0.848$ (M2) and $-0.798$ (M4). Walk and access time coefficients are consistently larger in magnitude than in-vehicle time, by approximately 28--31\%, indicating a higher behavioural penalty associated with out-of-vehicle time components. This finding is consistent with the established transport economics literature, which documents systematically higher valuations for walking, waiting, and access time relative to in-vehicle travel time, as evidenced by both early and recent meta-analyses of time multipliers \citep{wardman2004public, small2012valuation, wardman2026value}. For transit planning, this result implies that improvements to first and last-mile connectivity yield proportionally larger perceived benefits than equivalent reductions in line-haul travel time. The distance coefficients reveal a clear modal hierarchy where longer trips increase the relative attractiveness of car ($\beta_{D1} \approx 0.35$--$0.39$) and train ($\beta_{D3} \approx 0.25$--$0.31$), while each additional kilometre substantially reduces the likelihood of walking or cycling ($\beta_{D4} \approx -0.58$ to $-0.86$). In terms of the e-mobility options alternative that was introduced in the SP scenarios, the alternative-specific constant is not significant in either joint model ($-1.294$, $t = -1.57$ in M3; $-0.810$, $t = -0.35$ in M4), which suggests no inherent preference penalty toward e-scooters and shared electric bikes, and that the e-mobility options competes purely on its level-of-service attributes.
 
In both MXL specifications, the immigrant shift parameter ($\delta_{\text{MIG}}$) is positive and statistically significant, with estimates of $0.556$ in M2 (robust $t = 5.25$) and $0.525$ in M4 (robust $t = 3.58$). This shift reduces the magnitude of the in-vehicle travel time (IVTT) coefficient for immigrants, resulting in mean values of $-0.292$ (M2) and $-0.273$ (M4), compared to $-0.848$ and $-0.798$ for Canadian-born individuals. This corresponds to an approximate 66\% reduction in time sensitivity among immigrants. The estimated standard deviation of the random IVTT coefficient ($\sigma_T \approx 0.71$) is of similar magnitude to the population mean, which suggests that additional unobserved preference heterogeneity beyond the observed immigrant effect. 

In the MNL specifications, an additional immigrant-specific penalty for subway use is identified, with coefficients of $-0.708$ (M1) and $-0.653$ (M3). This indicates the presence of barriers to subway use beyond observed attributes such as travel time and cost, potentially reflecting factors such as wayfinding challenges or perceived safety. In the MXL specifications, this effect is no longer separately identified, as it is captured within the random parameter structure.

The integration effects are negative across both public transport and active modes. For public transport, the integration coefficient ($\beta_{I1}$) is estimated at $-0.179$ in M1 ($t = -9.03$), $-0.095$ in M2 ($t = -0.55$), $-0.213$ in M3 ($t = -10.18$), and $-0.106$ in M4 ($t = -0.73$). For active modes, the corresponding coefficients are $-0.259$ (M1) and $-0.273$ (M3). Statistical significance is strongest in the MNL specifications, while the MXL models retain the same directional effects with wider confidence intervals, consistent with random parameters capturing a substantial share of individual-level heterogeneity.

These estimates indicate that higher levels of integration are associated with a reduced likelihood of using both public transport and active modes. This pattern is consistent with the literature on immigrants travel behaviour, which documents a gradual convergence of travel patterns towards those of the host population over time, often accompanied by increased reliance on private vehicle use \citep{chatman2009immigrants, blumenberg2010getting, hu2021comparing, preston2023regionalization}.

To assess whether the integration gradient could be driven by income differences rather than embeddedness, a sensitivity analysis was conducted in which an income variable was introduced into the M1 specification alongside the integration index, interacted with both car and public transit utilities. The integration coefficient was unaffected by the inclusion of income: $\beta_{I1} = -0.179$ ($t = -9.02$) with income included, compared to $-0.179$ ($t = -9.03$) without. This stability is consistent with the model structure, in which the cost coefficient already captures price sensitivity and car availability is conditioned on at the trip level, absorbing the primary channels through which income influences mode choice. Within the immigrant, income shows no significant association with the social ($\rho = 0.11$, $p = 0.55$), civic ($\rho = 0.19$, $p = 0.29$), or health ($\rho = 0.23$, $p = 0.21$) sub-dimensions that together constitute 60\% of the index weight, which indicates that the integration index captures dimensions of embeddedness that are empirically distinguishable from income.
 
The estimated SP scale parameters are $\mu_{\text{SP}} = 0.298$ in M3 ($t = 5.51$) and $0.281$ in M4 ($t = 2.76$), imply that SP error variances are larger than those of RP. A distinctive feature of this study is the SP design, in which scenarios were generated dynamically within the mobile application based on participants’ observed trips. This approach enhances ecological validity by capturing realistic mode-switching decisions, but it may also introduce additional response variability. 

A key contributor to the observed scale differential is sample imbalance, with approximately 14{,}500 RP observations compared to 622 SP observations, corresponding to a ratio exceeding 23:1. To assess whether the low $\mu_{\text{SP}}$ reflects genuine hypothetical bias or is driven by this imbalance, the joint MNL model was re-estimated on a balanced subsample including only RP trips that triggered SP scenarios, while holding sociodemographic parameters fixed at their Model~3 estimates. Under this specification, $\mu_{\text{SP}}$ increases to $0.491$ ($t = 4.69$). This suggests that a substantial share of the apparent SP noise in the full-sample estimation is attributable to sample imbalance rather than inherent hypothetical bias.

Despite the differences in scale and error structure, the core parameter estimates remain stable across specifications. Several sociodemographic effects are consistently identified, including the student effect on train ($\beta_{St} \approx -1.55$), the full-time employment effect on subway ($\beta_F \approx -0.25$ to $-0.68$), and the cycling-friendliness effect on bicycle ($\beta_{Cy} \approx 0.71$). In the MXL specifications, the inclusion of random parameters captures a substantial share of individual-level heterogeneity, which reduces the need for explicit interaction terms and yields more parsimonious model formulations.

To evaluate the out-of-sample predictive performance and robustness of the model specifications, five-fold observation-level cross-validation was conducted and results are presented in Table~\ref{tab:modelfit}. The results show mean prediction accuracies ranging from 79.6\% to 82.0\%, with standard deviations below 1\% across all specifications, which indicates a stable predictive performance. The MNL specifications achieve marginally higher predictive accuracy than their MXL counterparts (82.0\% vs.\ 79.9\% for RP-only models; 81.8\% vs.\ 79.6\% for RP--SP models), consistent with the well-documented tendency of MXL models to exhibit slightly lower predictive accuracy when predictions are based on population-mean parameters rather than individual-specific preference estimates.

\begin{figure}[htbp]
\centering
\includegraphics[width=0.8\textwidth]{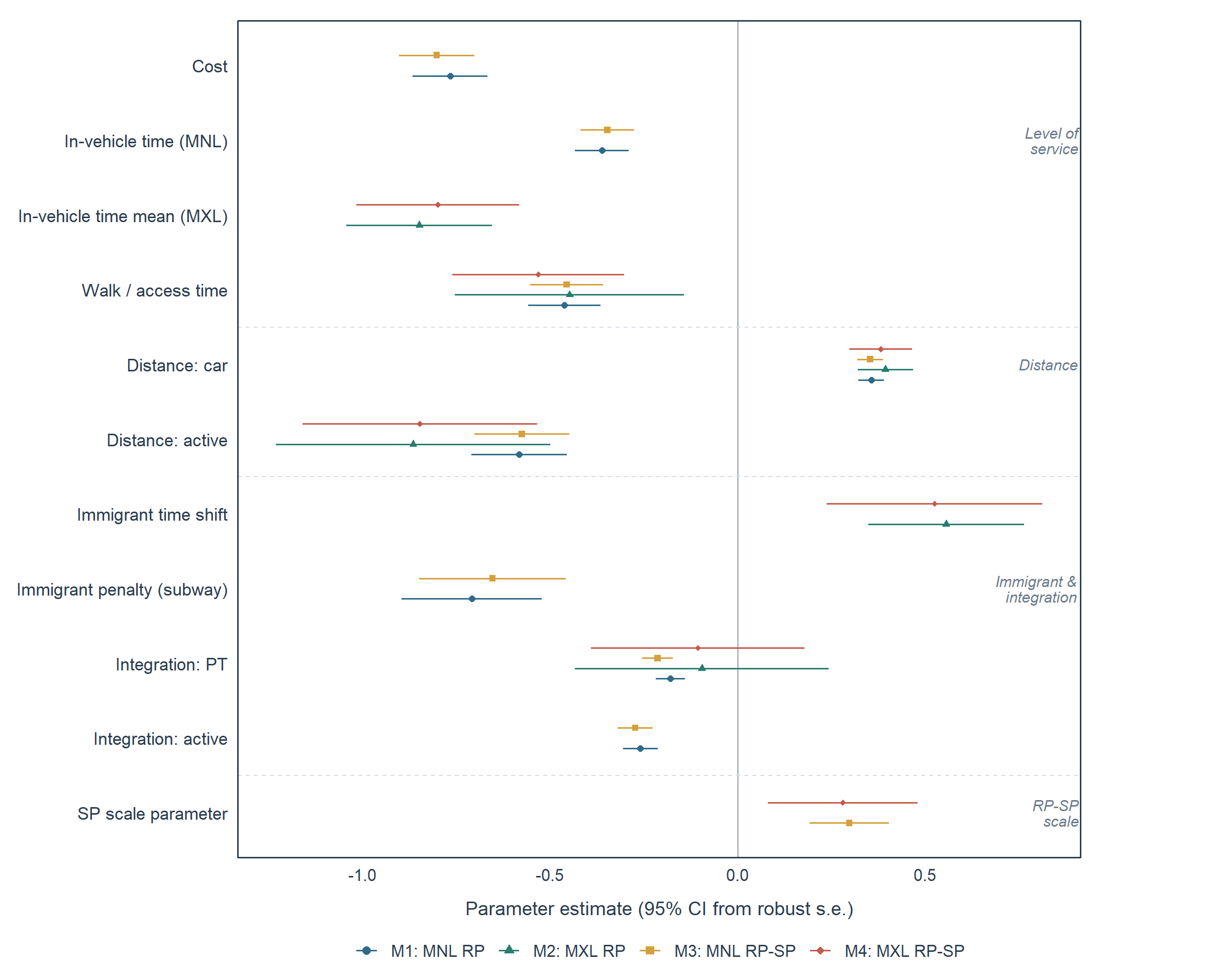}
\caption{Parameter estimates across all four model specifications with 95\% confidence intervals from robust standard errors, grouped by parameter class.}
\label{fig:forest}
\end{figure}

Parameter estimates are also stable across validation folds: $\beta_{I1}$ remains negative in all five folds for both M1 and M3, while $\delta_{\text{MIG}}$ is positive in all MXL folds. Figure~\ref{fig:forest} presents a forest plot of the key parameter estimates with 95\% confidence intervals across all four specifications, and Table~\ref{tab:stability} summarizes the cross-specification stability patterns.
 
\begin{table}[htbp]
\centering
\caption{Stability of key parameters across specifications}
\label{tab:stability}
\footnotesize
\begin{tabular}{l cccc l}
\toprule
\textbf{Parameter} & \textbf{M1} & \textbf{M2} & \textbf{M3} & \textbf{M4} & \textbf{Assessment} \\
\midrule
$\beta_C$ (Cost)            & $-$$^{***}$ & $-$$^{***}$ & $-$$^{***}$ & $-$$^{***}$ & Robust \\
$\beta_T$ (IVTT)            & $-$$^{***}$ & $-$$^{***}$ & $-$$^{***}$ & $-$$^{***}$ & Robust \\
$\delta_{\text{MIG}}$       & ---         & $+$$^{***}$ & ---         & $+$$^{***}$ & Robust \\
$\beta_{I1}$ (Integ. $\to$ PT) & $-$$^{***}$ & $-$ (n.s.) & $-$$^{***}$ & $-$ (n.s.) & MNL: robust; MXL: absorbed \\
$\beta_M$ (immigrant $\to$ Sub)  & $-$$^{***}$ & ---         & $-$$^{***}$ & --- & Robust in MNL \\
$\mu_{\text{SP}}$           & ---         & ---         & $+$$^{***}$ & $+$$^{***}$ & Robust \\
$\alpha_{\text{emob}}$      & ---         & ---         & n.s.        & n.s.       & No inherent penalty \\
immigrant/Can.-born VOT ratio & ---        & 0.34        & ---         & 0.34       & 66\% lower \\
\bottomrule
\end{tabular}
\end{table}

\subsection{Behavioural Insights and Economic Analysis}\label{sec:behavioural}

To provide an economic interpretation of the estimated preferences, Table~\ref{tab:vot} reports the implied values of travel time (VOT). The MNL specifications yield VOT estimates of \$26--28\,CAD/hr for in-vehicle travel time and \$34--36\,CAD/hr for walk and access time. These values lie between population-average appraisal parameters commonly used in Canadian practice and the higher estimates reported in discrete choice studies of time-sensitive travel contexts, positioning the results within the mid-range of empirically observed VOT values \citep{metrolinx2020bcm, xia2019estimating}.

The estimated premium for walk and access time relative to in-vehicle time indicates that out-of-vehicle time carries a higher behavioural penalty. This consistent with established findings in the transport literature. The magnitude of this premium is nevertheless more moderate than the generalized time weights commonly adopted in appraisal frameworks, which assign substantially higher penalties to walking and access components relative to in-vehicle travel time \citep{metrolinx2020bcm}.
 
\begin{table}[htbp]
\centering
\caption{Value of travel time}
\label{tab:vot}
\footnotesize
\begin{tabular}{llcc}
\toprule
& & \textbf{VOT$_{\text{IVTT}}$} & \textbf{VOT$_{\text{Walk}}$} \\
\textbf{Model} & \textbf{Population} & \textbf{(CAD/hr)} & \textbf{(CAD/hr)} \\
\midrule
\multicolumn{4}{l}{\textit{MNL models (fixed coefficients)}} \\
Model 1 (MNL RP)    & All & 28.3 & 36.1 \\
Model 3 (MNL RP-SP) & All & 26.0 & 34.1 \\[4pt]
\multicolumn{4}{l}{\textit{MXL models (population-mean coefficients)}} \\
Model 2 (MXL RP)    & Canadian-born   & \multicolumn{2}{c}{$\mu_T / \text{E}[\beta_C] \times 60$} \\
                     & immigrant        & \multicolumn{2}{c}{$(\mu_T + \delta_{\text{MIG}}) / \text{E}[\beta_C] \times 60$} \\
                     & Ratio (immigrant/Can.-born) & \multicolumn{2}{c}{0.34} \\[2pt]
Model 4 (MXL RP-SP) & Ratio (immigrant/Can.-born) & \multicolumn{2}{c}{0.34} \\
\bottomrule
\multicolumn{4}{p{11cm}}{\footnotesize MNL VOT: $\mu_{\text{SP}}$ cancels in the ratio. MXL ratio: $(\mu_T + \delta_{\text{MIG}}) / \mu_T$. Walk/access time valued 28--31\% above IVTT.}
\end{tabular}
\end{table}
 
The estimated immigrant-to-Canadian-born VOT ratio of approximately 0.34 in both MXL specifications suggests that the sampled immigrants place a lower marginal value on travel time than Canadian-born residents. While prior studies have documented differences in travel behaviour between immigrant and native-born populations, these have typically relied on cross-sectional specifications with binary immigrant indicators \citep{chatman2009immigrants, blumenberg2010getting}. The present results instead exploit panel structure and variation in integration, allowing for a more detailed identification of heterogeneity in time--cost trade-offs
This finding is consistent with evidence from Canadian contexts that suggests recent immigrants exhibit distinct travel preferences that evolve over time and gradually converge toward native-born patterns \citep{xu2018transportation}. Several mechanisms may contribute to this differential, including lower opportunity costs of time associated with underemployment, persistence of travel norms from countries of origin, and more limited access to faster travel alternatives, which may render longer travel times less discretionary.

\begin{figure}[htbp]
\centering
\begin{subfigure}[b]{0.48\textwidth}\centering
\includegraphics[width=\textwidth]{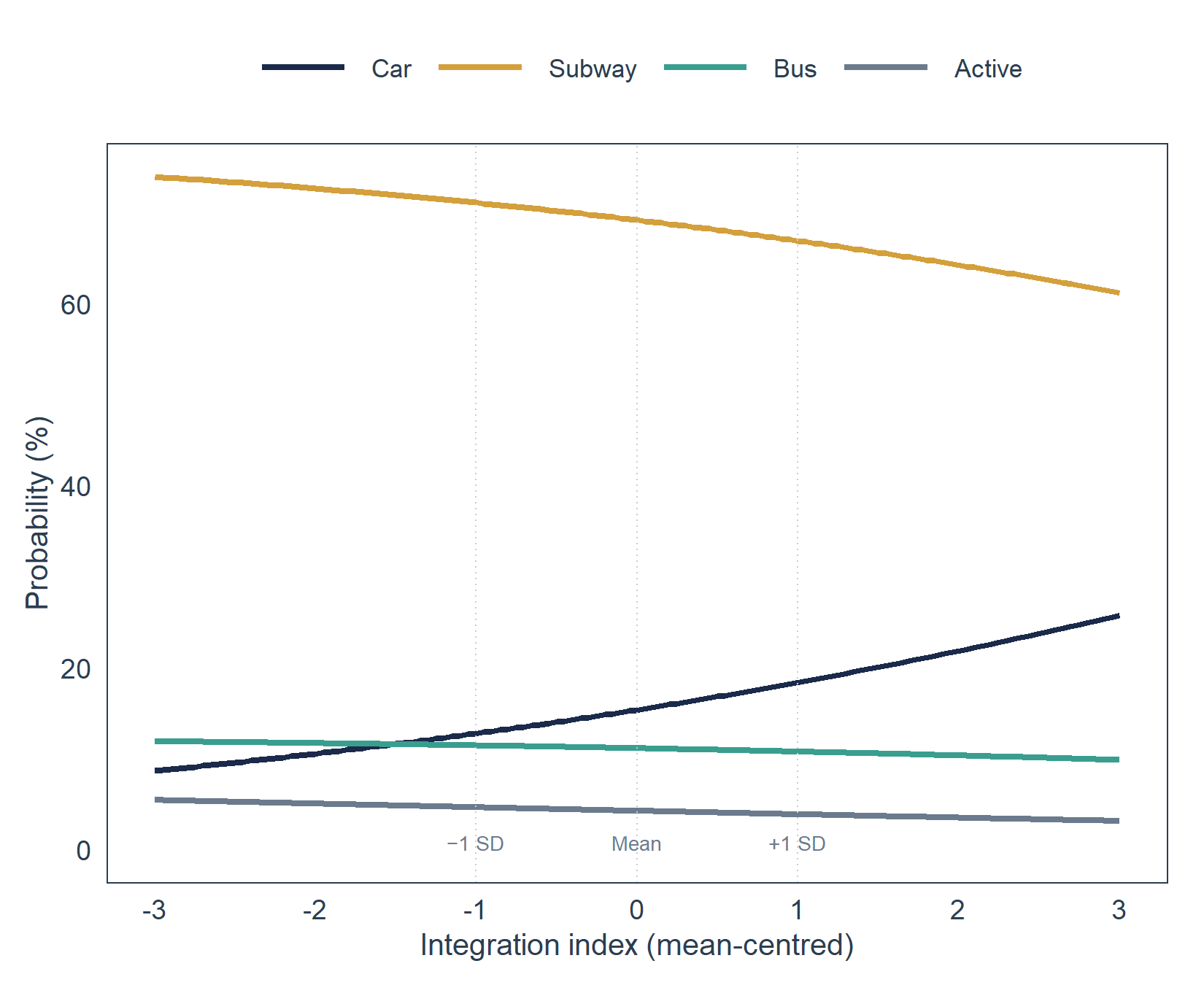}
\caption{All mode probabilities}
\label{fig:integ_all}
\end{subfigure}\hfill
\begin{subfigure}[b]{0.48\textwidth}\centering
\includegraphics[width=\textwidth]{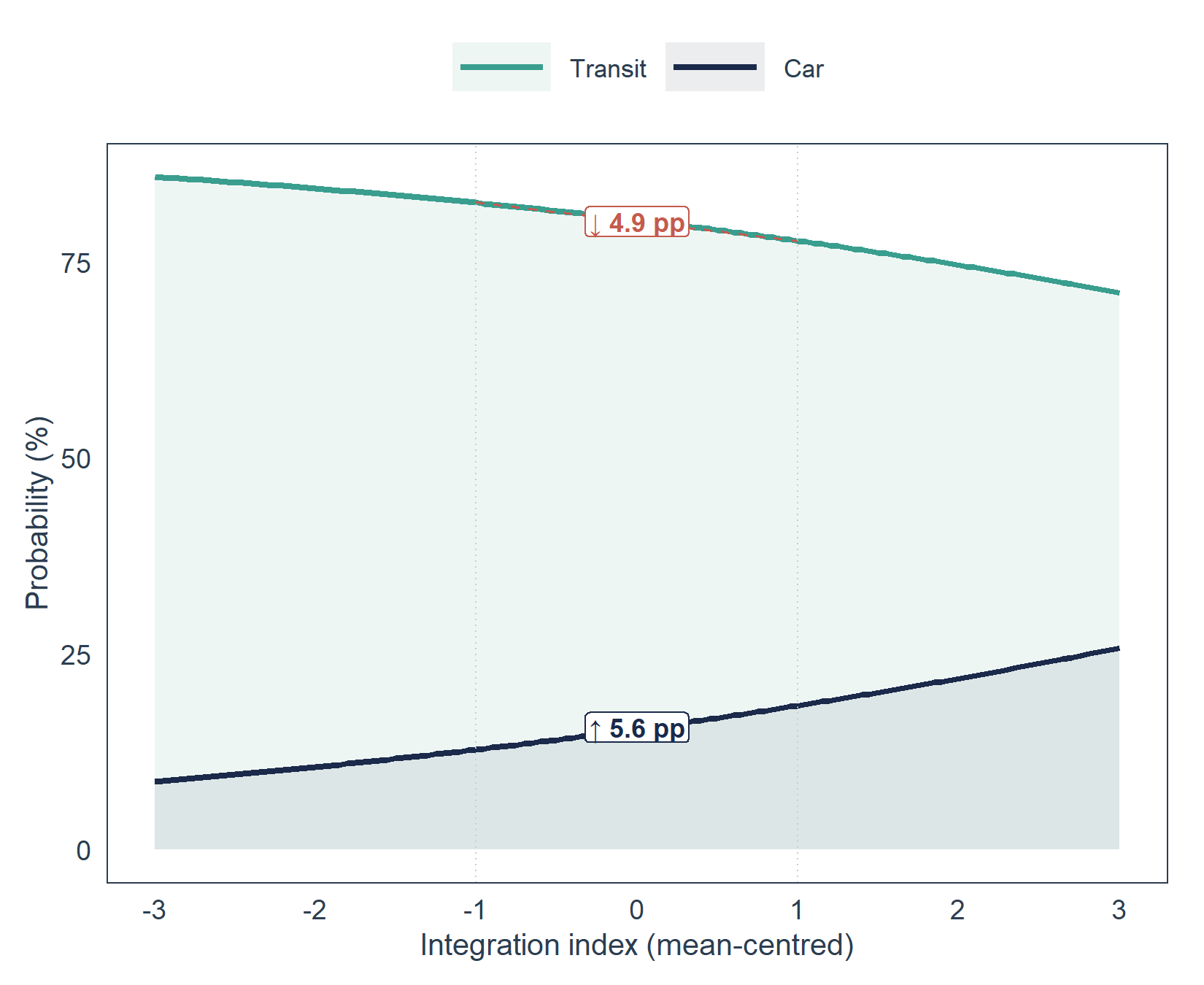}
\caption{Car vs transit}
\label{fig:integ_carvstransit}
\end{subfigure}
\caption{Integration effect on mode choice probabilities (Model~3, representative 10~km work commute): \subref{fig:integ_all} all mode probabilities and \subref{fig:integ_carvstransit} car vs.\ combined transit with percentage-point gap annotations.}
\label{fig:integration_effect}
\end{figure}

To further interpret the behavioural implications of the estimated model, we examine how integration translates into changes in mode choice probabilities under representative travel conditions. Figure~\ref{fig:integration_effect} presents the integration gradient for a representative immigrant making a 10~km work commute, using Model~3 parameters.Moving from one standard deviation below the mean to one above is associated with a reduction in combined transit probability of approximately 5 percentage points while increasing car probability by a similar margin. The effect is continuous and near-linear over the observed range of integration scores, which suggests that integration may erode transit use gradually rather than through a discrete threshold.

Building on the behavioural patterns illustrated in Figure~\ref{fig:integration_effect}, we translate the estimated parameters into policy-relevant outcomes by simulating mode choice probabilities under counterfactual scenarios using Model~3. The MNL structure of M3 yields closed-form choice probabilities for any hypothetical attribute profile, avoiding the additional simulation layer required by MXL specifications with random parameters. M3 is also the most comprehensive MNL specification, incorporating both RP and SP data and the full alternative set, including e-mobility options.
 
We consider a representative immigrant making a 10~km work commute and examine two intervention levers: transit fare reduction and access time reduction, selected as the primary policy instruments available to transit agencies. For each scenario, transit choice probabilities are computed at three integration levels to assess whether intervention effectiveness varies across the integration spectrum.
 
Table~\ref{tab:shift_fare} presents the fare reduction scenario. Eliminating transit fares increases transit probability by approximately 3 to 4 percentage points, depending on integration level. The effect is modest, and reflects the linear role of cost in the utility function and the relatively small fare differential between car and transit for a 10~km trip. Even under a zero-fare scenario, the difference between low- and high-integration immigrants persists at over 4 percentage points.
 
\begin{table}[htbp]
\centering
\caption{Transit probability under fare reduction scenarios (\%), Model~3}
\label{tab:shift_fare}
\footnotesize
\begin{tabular}{lccc}
\toprule
\textbf{Fare (CAD)} & \textbf{Low integ.} & \textbf{Avg integ.} & \textbf{High integ.} \\
& ($-1$ SD) & (Mean) & ($+1$ SD) \\
\midrule
\$3.25 (current) & 82.6 & 80.4 & 77.7 \\
\$2.50           & 83.4 & 81.3 & 78.7 \\
\$1.50           & 84.5 & 82.5 & 80.1 \\
\$0.50           & 85.5 & 83.6 & 81.3 \\
\$0.00 (free)    & 86.0 & 84.1 & 81.9 \\
\midrule
\textit{Gain (pp)} & \textit{+3.4} & \textit{+3.7} & \textit{+4.2} \\
\bottomrule
\multicolumn{4}{p{10cm}}{\footnotesize Representative immigrant, 10~km work commute, full-time employed. Integration measured in standard deviations from the mean.}
\end{tabular}
\end{table}
 
Table~\ref{tab:shift_access} presents the access time reduction scenario, starting from a 15-minute baseline. In contrast to the fare intervention, reducing walk-to-stop time produces substantially larger gains in transit use. Eliminating access time increases transit probability by approximately 10 to 12 percentage points across integration levels, roughly three times the effect of complete fare elimination. This result is consistent with the higher behavioural penalty associated with out-of-vehicle time: reductions in access time simultaneously improve the perceived utility of all transit alternatives.
 
\begin{table}[htbp]
\centering
\caption{Transit probability under access time reduction scenarios (\%), Model~3}
\label{tab:shift_access}
\small
\begin{tabular}{lccc}
\toprule
\textbf{Access time (min)} & \textbf{Low integ.} & \textbf{Avg integ.} & \textbf{High integ.} \\
& ($-1$ SD) & (Mean) & ($+1$ SD) \\
\midrule
15 (baseline) & 77.7 & 75.1 & 72.0 \\
10            & 81.4 & 79.1 & 76.3 \\
5             & 84.6 & 82.6 & 80.2 \\
2             & 86.3 & 84.5 & 82.3 \\
0             & 87.4 & 85.6 & 83.6 \\
\midrule
\textit{Gain (pp)} & \textit{+9.7} & \textit{+10.5} & \textit{+11.6} \\
\bottomrule
\multicolumn{4}{p{10cm}}{\footnotesize Same representative immigrant. Fare held at current \$3.25 CAD.}
\end{tabular}
\end{table}
 
The effectiveness of this lever is particularly pronounced for highly integrated immigrants, who exhibit the largest gains in transit probability. This reflects their lower baseline propensity to use transit and greater responsiveness to improvements in access, which places them closer to the margin between transit and car use.
 
Through this analysis, two patterns emerge. First, access time reduction is consistently the more effective lever across all integration levels, highlighting the importance of first- and last-mile connectivity relative to fare-based interventions. Second, the integration gap persists under all scenarios: even under the combined effect of zero fares and zero access time, a non-negligible difference remains between low- and high-integration immigrants. This suggests that the shift toward car-oriented preferences associated with integration operates through channels beyond cost and access, potentially including habit formation, social norms, and evolving perceptions of convenience and status.
 
Overall, the results suggest a behavioural pattern in which integration is associated with a gradual shift away from public transport toward car-oriented travel, driven by differences in time-cost trade-offs. Policy interventions such as fare reductions and access time improvements can partially moderate this shift, but their effects are heterogeneous and insufficient to eliminate the integration gap. Improvements in first- and last-mile connectivity are consistently more effective than fare-based measures, yet even substantial changes in service attributes do not fully offset the underlying preference shift. These patterns suggest that the evolution of travel behaviour with integration may extend beyond observable attributes, and that conventional policy instruments alone may not be sufficient to counteract increasing car reliance among integrating populations.

\section{Discussion}\label{sec:discussion}
 
Several directions for sustainable mobility planning in immigrant-receiving cities merit consideration: retaining early transit users, prioritizing infrastructure over fare subsidies, leveraging the settlement period as a behavioural window, rethinking transport appraisal methods, and improving transit legibility for immigrants.
 
\subsection{Retaining Immigrants as Public Transit Users}
 
The central policy challenge identified by this study is that integration into Canadian society erodes public transit use. Each standard deviation increase in integration reduces transit probability by approximately 5 percentage points (Figure~\ref{fig:integration_effect}), and this gradient persists even under the most generous fare and access time scenarios. As immigrants settle, acquire driver's licences, purchase vehicles, and adopt the travel norms of their Canadian-born neighbours, they may progressively shift from transit to car, a transition that could undermine both ridership and emission-reduction targets.
 
This integration-driven car adoption pathway is not inevitable. The stated preference analysis (Figure~\ref{fig:sp}) suggests that the sampled immigrants appear more willing than Canadian-born participants to consider alternative modes across all trip categories. The estimated lower time sensitivity ($\delta_{\text{MIG}} > 0$, 66\% reduction) suggests that immigrants are less deterred by the travel time disadvantage of transit relative to car. These characteristics create a window during the early settlement period in which transit habits can be reinforced before car-oriented patterns crystallize.
 
Practical retention strategies include time-limited mobility support packages for immigrants, combining free transit passes, bike-share memberships, micro-mobility credits, and multilingual trip-planning assistance during the first one to two years of settlement. The finding that e-mobility options faces no inherent preference penalty suggests that new mobility services can be incorporated into these packages without requiring attitudinal shifts, provided the cost and access attributes are competitive. Settlement agencies and transit operators have a shared interest in coordinating such programmes, as sustained transit ridership among immigrant populations supports both integration outcomes and public transport financial viability.
 
\subsection{Access Time Over Fare: Prioritizing Infrastructure}
 
The modal shift simulations reveal that access time reduction is approximately three times more effective than fare elimination in shifting car commuters to transit (11.6~pp vs.\ 4.2~pp for high-integration immigrants). This finding, combined with the 28--31\% premium that all respondents place on walk/access time over in-vehicle time, suggests that first/last-mile infrastructure investment may be a more effective lever than fare subsidies for encouraging sustainable mode shift.
 
For immigrant-dense suburban neighbourhoods, where residential choices may be constrained by housing affordability rather than transit proximity, this means investing in feeder bus routes, especially in the form of on-demand transit, protected pedestrian and cycling connections to rapid transit stations, and real-time multimodal journey planning in multiple languages. Such investments could reduce the access time barrier that the models suggest disproportionately affects less-integrated populations living further from transit corridors.
 
The persistence of the integration gap under all simulated scenarios suggests that infrastructure improvements alone may not be sufficient. The preference shift toward car dependence that accompanies integration operates through habit formation, social norm adoption, and changing perceptions of status and convenience, mechanisms that require complementary soft interventions including workplace travel plans, community-based mobility coaching, and targeted marketing of transit benefits to settled immigrant communities.
 
\subsection{Leveraging Migration as a Mobility Biography Disruption}
 
International migration represents one of the most significant disruptions to an individual's mobility biography. Arrival in a new country forces a complete reassessment of travel options: familiar routes, modes, and habits no longer apply. The travel behaviour literature recognizes such life events, including residential relocation, and job change as windows of opportunity for sustainable mode adoption, when habitual behaviour is temporarily unfrozen and individuals are receptive to new options.
 
The estimated models suggest this window may be quantifiable. Immigrants exhibit a 66\% reduction in time sensitivity relative to Canadian-born residents, a higher willingness to explore alternatives in stated preference scenarios, and a transit-oriented baseline that persists through the early stages of integration. As integration progresses, these characteristics converge toward Canadian-born norms. This suggests that interventions delivered during the early stages of settlement, when travel habits are still forming, could have larger and more lasting effects compared to the same interventions delivered to established residents with entrenched car habits.
 
This aligns with emerging calls for just transition pathways that link emission reduction with social inclusion. immigrant populations, which constitute over 23\% of Canada's population and are projected to grow, may represent an underexplored opportunity for sustainable mobility. Coupling behavioural evidence from studies such as this one with integration-sensitive policy design could contribute toward both climate targets and more equitable, resilient transport systems.
 
\subsection{Implications for Transport Appraisal}
 
The estimated immigrant/Canadian-born value-of-travel-time ratio of 0.34 could have implications for cost-benefit analysis (CBA) of transit infrastructure projects. Standard transport appraisal methods typically apply a single population-average VOT drawn from national or regional guidelines. In corridors with high concentrations of immigrants, this practice systematically overstates the user benefits of time-saving improvements while understating the retention benefits of maintaining existing transit service quality.
 
A more nuanced appraisal framework would incorporate population-segmented VOTs. In corridors where immigrants constitute a substantial share of ridership, the case for transit investment rests less on conventional time-savings benefits and more on ridership retention, social inclusion, and the avoidance of long-term car dependency. Agencies conducting CBA for transit projects in cities such as Toronto, Montr\'eal, and Vancouver, where immigrant populations are concentrated along specific transit corridors, could benefit from considering whether differential time valuations of the kind estimated here apply to their ridership base.
 
\subsection{Improving Transit Legibility and Wayfinding}
 
The immigrant-specific subway penalty identified in the MNL models which operates independently of the generic time sensitivity differential, may point to barriers beyond cost and travel time. Underground rapid transit systems can present navigational challenges for immigrants such as complex station layouts, transfer procedures, fare payment systems, and service disruption information are typically communicated in ways that assume local familiarity. These barriers may be compounded by safety perceptions that differ from those of Canadian-born riders, particularly in unfamiliar subterranean environments.
 
Targeted design interventions can address these barriers directly. Simplified network maps with intuitive colour coding, pictographic wayfinding systems that reduce reliance on text, and mobile journey-planning applications with step-by-step guidance would reduce the cognitive burden of using rapid transit. Several transit agencies internationally have implemented such measures with demonstrated ridership gains among immigrant populations. The fact that this subway penalty does not appear in the MXL specifications, where it is absorbed by the random parameters, suggests that it reflects a distributional characteristic of immigrant preferences rather than a uniform barrier, which means that targeted interventions need not be universal but can be concentrated at stations and corridors serving immigrant-dense neighbourhoods.
 
\section{Conclusion}\label{sec:conclusion}
 
This study investigated the relationship between immigrant status, integration, and transportation mode choice in Canadian cities using a methodological framework that combines four elements: a continuous multidimensional integration index, mixed logit models with systematic immigrant-linked taste heterogeneity, joint revealed-stated preference estimation, and a GPS-based mobile app panel dataset of more than 80,000 trip observations and 622 SP scenarios from 100 participants.
 
Three patterns emerge from the estimated models. First, the sampled immigrants appear to be approximately 66\% less sensitive to in-vehicle travel time than Canadian-born residents, translating to a value-of-travel-time ratio of 0.34 in both MXL specifications. This differential, captured through the systematic heterogeneity parameter in the mixed logit models. Second, integration into Canadian society is associated with a progressive shift from public transit toward car dependence: each standard deviation increase in the integration index reduces transit probability by approximately 5 percentage points, an effect that persists even under the most generous fare and access time scenarios. Third, e-mobility options alternatives face no inherent preference penalty, competing purely on their level-of-service attributes.
 
The modal shift analysis demonstrates that access time reductions are approximately three times more effective than fare elimination in attracting car commuters to transit, and that the integration-driven car adoption gradient cannot be fully closed by service improvements alone. These patterns suggest that the early settlement period may represent a window for sustainable mobility intervention, when immigrants' lower time sensitivity, higher mode flexibility, and transit-oriented baseline offer a receptive context for habit formation that diminishes as integration progresses.
 
The 2$\times$2 model comparison framework (MNL vs.\ MXL, RP vs.\ joint RP-SP) confirms that these results are not artefacts of particular modelling choices. The core findings hold across model structures and data sources, with five-fold observation-level cross-validation yielding mean prediction accuracy of 80--82\% with standard deviations below 1\%. The continuous integration index, applied to both immigrant and Canadian-born participants, captures a dimension of socioeconomic embeddedness that the conventional binary immigrant/Canadian-born distinction cannot represent. 
 
For transit agencies and policymakers in cities receiving immigrant populations, the findings suggest five directions that merit consideration: early-stage mobility packages that leverage immigrants' openness to alternatives before car habits form; infrastructure investment in first/last-mile access rather than fare subsidies; population-segmented transport appraisal that accounts for differential time valuations; improved transit legibility through simplified navigation; and integration-aware demand management that recognizes the settlement period as a critical window for sustainable travel behaviour formation.
 
%\subsection*{Limitations and Future Research}
 
Several limitations should be acknowledged. The sample of 100 individuals, while generating a large volume of trip observations through the multi-day panel design, limits the number of unique sociodemographic profiles available for estimating taste heterogeneity. The mixed logit random parameters are identified from the full sample, but the precision of the estimated distributions, particularly for the immigrant shift $\delta_{\text{MIG}}$, would benefit from larger samples with greater demographic diversity. Nonetheless, the core behavioural parameters, including the immigrant time sensitivity differential and the integration gradient, are estimated with strong statistical significance and confirmed through five-fold cross-validation, which indicates that the panel depth compensates adequately for the limited cross-sectional breadth for the purposes of this study.
The sample skews male (70\% overall) and is concentrated in central Toronto and Montreal, where transit infrastructure is well-developed. Generalizability to suburban contexts, smaller cities, or regions with limited transit provision may be restricted. The higher walking and cycling propensities documented among immigrants may not hold in car-dependent suburban environments where active travel infrastructure is sparse.

The first generation immigrants subsample encompasses participants from diverse countries of origin and cultural backgrounds. While immigrants from different regions may carry distinct travel norms and preferences, the analysis treats them as a single group defined by the shared structural condition of being foreign-born in Canada. The integration index partially addresses this by capturing individual-level variation in socioeconomic embeddedness regardless of origin. Nevertheless, the sample size does not permit disaggregation by country or region of origin, and the behavioural patterns identified here should be interpreted as reflecting the common experience of settlement rather than origin-specific cultural effects. 

Future research with larger and more diverse immigrant samples could investigate whether the time sensitivity differential and integration gradient vary by source country or immigration pathway.
The lognormal cost distribution in the MXL specifications, while ensuring the theoretically correct negative sign, produces heavy-tailed posteriors that preclude reliable individual-level VOT estimation via conditional parameters. Future work could explore bounded distributions (e.g.\ triangular or truncated normal) or willingness-to-pay space models that directly estimate the VOT distribution.
Cycling accounts for only 2.3\% of observed trips. Targeted oversampling of cyclists in future surveys would enable more reliable estimation of cycling-specific parameters.
Finally, latent class models were explored during specification development but proved unviable at this sample size, with class collapse and unreliable estimates across all attempted configurations. Larger number of participants may enable latent class or hybrid choice models that capture discrete population segments alongside the continuous heterogeneity modelled here. Investigating the use of generative AI for fusing data from conventional travel surveys such as the Transportation Tomorrow Survey in Toronto and enquêtes origine-destination in Montr\'eal (high breadth) with BDMobility app based surveys (high depth) is another future direction that can greatly increase our ability to better represent immigrant behaviour in travel demand modelling.

\section*{Acknowledgements}
This research is funded by the Canada First Research Excellence Fund under the Bridging Divides program and the Canada Research Chair in Disruptive Transportation Technologies and Services (CRC–2021–0048). The authors thank Bara Rababah, Henrique De Freitas Serra, Isaac Otchere, and Thomas Zhao for their support in data collection and participant recruitment.

\setlength{\bibsep}{0.0pt} 
\bibliographystyle{abbrvnat}
\singlespacing
\bibliography{References.bib}
\end{document}